\PassOptionsToPackage{table}{xcolor}
\PassOptionsToPackage{bookmarks=true,breaklinks=true,letterpaper=true,colorlinks,citecolor=blue,linkcolor=blue,urlcolor=blue}{hyperref}

\PassOptionsToPackage{dvipsnames}{xcolor}

\documentclass[sigplan, 10pt, nonacm]{acmart}
\renewcommand\footnotetextcopyrightpermission[1]{}
\pdfoutput=1  %

\usepackage{enumitem}
\usepackage{soul}
\usepackage{microtype}
\usepackage{multicol}
\usepackage{subcaption}
\usepackage[utf8]{inputenc}
\usepackage{xcolor}
\usepackage{rotating}
\usepackage{makecell}
\usepackage{titlesec}

\soulregister{\us}{7}
\soulregister{\mprotect}{7}
\soulregister{\xname}{7}
\soulregister{\memcpy}{7}
\soulregister{\todo}{7}

\makeatletter                   
\def\mdseries@tt{m}             
\makeatother                    

\usepackage[draft=true]{minted} 
\usepackage{color}
\usepackage{textgreek}

\usepackage{amssymb}
\usepackage{lipsum}
\usepackage[final]{pdfpages}
\usepackage{setspace}
\usepackage{epsf}
\usepackage{epsfig}
\usepackage{graphicx}
\usepackage{algorithmic}
\usepackage{amsmath}
\usepackage{letltxmacro}

\usepackage{enumitem}
\usepackage{tikz}
\usepackage{tightenum}
\usepackage{xspace}

\usepackage{url}
\usepackage{multirow}

  \renewenvironment{thebibliography}[1]{%
    \begin{oldthebibliography}{#1}%
      \setlength{\parskip}{0ex}%
      \setlength{\itemsep}{0ex}%
  }%
  {%
    \end{oldthebibliography}%
  }

\newwrite\arxivdeps
\immediate\openout\arxivdeps=\jobname-arxivdeps.log

\newcommand\verifymarkedforarxivfile[1]{%
\ifdefined\arxivbuild
\else
\IfFileExists{#1}%
{}%
{\GenericWarning{Marked file (#1) for inclusion in arxiv build does not exist}}
\fi%
}

\newcommand\markforarxiv[1]{%
\verifymarkedforarxivfile{#1}%
\write\arxivdeps{IncludeInArxiv: #1}%
}

\DeclareUrlCommand\UScore{\urlstyle{rm}}

\LetLtxMacro\oldincludegraphics\includegraphics
\renewcommand{\includegraphics}[2][]{%
\markforarxiv{#2}%
\oldincludegraphics[#1]{#2}}

\LetLtxMacro\oldincludepdf\includepdf
\renewcommand{\includepdf}[2][]{%
\markforarxiv{#2}%
\oldincludepdf[#1]{#2}}

\def\nfigure[#1,#2,#3]{
\begin{figure}
\vspace*{0mm}
\begin{center}

\includegraphics[width=\columnwidth]{#1} 
\caption[]{#2
} \label{#3}

\end{center}
\end{figure}}

\def\cfigure[#1,#2,#3]{
\begin{figure}
\vspace*{0mm}
\begin{center}

\includegraphics[width=3in]{#1} 
\caption[]{#2
} \label{#3}

\end{center}
\end{figure}}

\def\cfigurefour[#1,#2,#3]{
\begin{figure}
\vspace*{0mm}
\begin{center}

\includegraphics[width=4in]{#1} 
\vspace*{-3mm}\caption[]{#2
} \label{#3}
 
\vspace*{-5mm}
\end{center}
\end{figure}}

\def\cfiguretemp[#1,#2,#3]{
\begin{figure}
\vspace*{0mm}
\begin{center}

\includegraphics[width=3.5in]{#1} 
\vspace*{-3mm}\caption[]{#2
} \label{#3}
 
\vspace*{-5mm}
\end{center}
\vspace*{-2mm}
\end{figure}}

\def\wfigure[#1,#2,#3]{
\begin{figure*}
\vspace*{0mm}
\begin{center}
 \includegraphics[width=\textwidth]{#1} 
 \vspace*{-3mm}\caption[]{#2
} \label{#3}
 
\end{center}
\end{figure*}}

\def\threefigure[#1,#2,#3,#4,#5]{
\begin{figure*}
\vspace*{0mm}
\begin{center}

\begin{tabular}{ccc}
\includegraphics[width=2in]{#1} & \includegraphics[width=2in]{#2} &  \includegraphics[width=2in]{#3} \\
(a) & (b) & (c) \\
\end{tabular}
\vspace*{-3mm}\caption[]{#4
} \label{#5}

\vspace*{-5mm}
\end{center}
\vspace*{-2mm}
\end{figure*}}

\def\dcfigure[#1,#2,#3,#4,#5,#6]{
{
\begin{figure*}
\begin{center}
\begin{minipage}[c]{\columnwidth}{
\includegraphics[width=\columnwidth]{#1} 
\vspace*{0mm}\caption[]{#2} \label{#3} \
}\end{minipage}\hspace*{\columnsep}\
\begin{minipage}[c]{\columnwidth}{
\includegraphics[width=\columnwidth]{#4} 
\vspace*{0mm}\caption[]{#5}\label{#6} \
}\end{minipage}
\end{center}
\end{figure*}
}
}

\def\scfigure[#1,#2,#3]{
{
\begin{figure*}
\begin{center}
\begin{minipage}[c]{3.5in}{
\includegraphics[width=3.5in]{#1} 
}\end{minipage}
\caption[]{#2} \label{#3} \
\end{center}
\end{figure*}
}
}

\def\tableByTable[#1,#2,#3,#4,#5,#6]{
{
\begin{table*}
\begin{center}
\begin{minipage}[c]{3in}{
\centering
{#1}
\vspace*{0mm}\tabcaption[]{#2}\label{#3} \
}\end{minipage}\hspace*{\columnsep}\
\begin{minipage}[c]{3in}{
\centering
{#4}
\vspace*{0mm}\tabcaption[]{#5}\label{#6} \
}\end{minipage}
\end{center}
\end{table*}
}
}

\def\figureByTable[#1,#2,#3,#4,#5,#6]{
{
\begin{figure*}
\begin{center}
\begin{minipage}[c]{3in}{
\centering
\includegraphics[width=\textwidth]{#1}
\vspace*{0mm}\figcaption[]{#2} \label{#3} \
}\end{minipage}\hspace*{\columnsep}\
\begin{minipage}[c]{3.3in}{
\centering
{#4}
\vspace*{0mm}\tabcaption[]{#5}\label{#6} \
}\end{minipage}
\end{center}
\end{figure*}
}
}

\def\tableByFigure[#1,#2,#3,#4,#5,#6]{
{
\begin{figure*}
\begin{center}
\begin{minipage}[c]{4.3in}{
\centering
{#1}
\vspace*{0mm}\tabcaption[]{#2} \label{#3} \
}\end{minipage}\hspace*{\columnsep}\
\begin{minipage}[c]{2.2in}{
\centering
\includegraphics[width=\textwidth]{#4}
\vspace*{-0.35in}\caption[]{#5}\label{#6} \
}\end{minipage}
\end{center}
\end{figure*}
}
}

\def\doublecfigure[#1,#2,#3,#4]{
{
\begin{figure}
\begin{center}
\begin{minipage}[c]{1.5in}{
\begin{center}
\includegraphics[width=1.5in]{#1}
\end{center}
}\end{minipage}\hspace*{1em}\
\begin{minipage}[c]{1.5in}{
\begin{center}
\includegraphics[width=1.5in]{#2}
\end{center}
}\end{minipage}
\vspace*{0mm}\caption[]{#3} \label{#4} \
\end{center}
\end{figure}
}
}

\def\qcfigure[#1,#2,#3,#4,#5,#6]{
{
\begin{figure*}
\vspace*{0.2in}\
\begin{center}
\begin{minipage}[c]{3in}{
\includegraphics[width=3in]{#1} 
\vspace*{-3mm}
}
\end{minipage}\hspace*{0.5in}\
\begin{minipage}[c]{3in}{
\includegraphics[width=3in]{#2} 
\vspace*{-3mm}
}\end{minipage}

\begin{minipage}[c]{3in}{
\includegraphics[width=3in]{#3} 
\vspace*{-3mm}
}
\end{minipage}\hspace*{0.5in}\
\begin{minipage}[c]{3in}{
\includegraphics[width=3in]{#4} 
\vspace*{-3mm}
}\end{minipage}
\end{center}
\caption[]{#5}\label{#6}
\end{figure*}
}
}

\def\twfigure[#1,#2,#3,#4,#5]{
{
\begin{figure*}
\vspace*{0.2in}\
\begin{center}
\begin{minipage}[c]{6.5in}{
\includegraphics[width=6.5in]{#1} 
\vspace*{-3mm}
}
\end{minipage}

\begin{minipage}[c]{6.5in}{
\includegraphics[width=6.5in]{#2} 
\vspace*{-3mm}
}\end{minipage}

\begin{minipage}[c]{6.5in}{
\includegraphics[width=6.5in]{#3} 
\vspace*{-3mm}
}
\end{minipage}
\end{center}
\caption[]{#4}\label{#5}
\end{figure*}
}
}

\def\dwfigure[#1,#2,#3,#4]{
{
\begin{figure*}
\vspace*{0.2in}\
\begin{center}
\begin{minipage}[c]{6.5in}{
\includegraphics[width=6.5in]{#1} 
\vspace*{-3mm}
}
\end{minipage}

\begin{minipage}[c]{6.5in}{
\includegraphics[width=6.5in]{#2} 
\vspace*{-3mm}
}\end{minipage}

\end{center}
\caption[]{#3}\label{#4}
\end{figure*}
}
}

\def\dssfigure[#1,#2,#3,#4,#5,#6]{
{
\begin{figure*}
\vspace*{0.2in}\
\begin{center}
\begin{minipage}[c]{4in}{
\includegraphics[width=4in]{#1}
\vspace*{-3mm}\caption[]{#2} \label{#3} \
}\end{minipage}\hspace*{0.5in}\
\begin{minipage}[c]{2in}{
\includegraphics[width=2in]{#4}
\vspace*{-3mm}\caption[]{#5}\label{#6} \
}\end{minipage}
\end{center}
\vspace*{-0.4in}\
\end{figure*}
}
}

\def\dsfigure[#1,#2,#3,#4,#5,#6]{
{
\begin{figure*}
\vspace*{0.2in}\
\begin{center}
\begin{minipage}[c]{3in}{
\includegraphics[width=3in]{#1}
\vspace*{-3mm}\caption[]{#2} \label{#3} \
}\end{minipage}\hspace*{0.5in}\
\begin{minipage}[c]{3in}{
\hspace*{0.5in}\
\includegraphics[height=3in]{#4}
\vspace*{-3mm}\caption[]{#5}\label{#6} \
}\end{minipage}
\end{center}
\vspace*{-0.4in}\
\end{figure*}
}
}

\def\dsyfigure[#1,#2,#3,#4,#5,#6]{
{
\begin{figure*}
\vspace*{0.2in}\
\begin{center}
\begin{minipage}[c]{2.5in}{
\includegraphics[height=2.5in]{#1}
\vspace*{-3mm}\caption[]{#2} \label{#3} \
}\end{minipage}\hspace*{0.5in}\
\begin{minipage}[c]{2.5in}{
\includegraphics[height=2.5in]{#4}
\vspace*{-3mm}\caption[]{#5}\label{#6} \
}\end{minipage}
\end{center}
\vspace*{-0.4in}\
\end{figure*}
}
}

\def\dyfigure[#1,#2,#3,#4,#5,#6]{
{
\begin{figure*}
\vspace*{0.2in}\
\begin{center}
\begin{minipage}[c]{3in}{
\includegraphics[height=3in]{#1} 
\vspace*{-3mm}\caption[]{#2} \label{#3} \
}\end{minipage}\hspace*{0.5in}\
\begin{minipage}[c]{3in}{
\includegraphics[height=3in]{#4} 
\vspace*{-3mm}\caption[]{#5}\label{#6} \
}\end{minipage}
\end{center}
\vspace*{-0.4in}\
\end{figure*}
}
}

\def\dyoldfigure[#1,#2,#3,#4,#5,#6]{
{
\begin{figure*}
\vspace*{0.2in}\
\begin{center}
\begin{minipage}[c]{3in}{
\epsfysize=2.0in\
\hspace{0.5in}\
\epsfbox{#1}
\vspace*{-3mm}\caption[]{#2} \label{#3} \
}\end{minipage}\hspace*{0.25in}\
\begin{minipage}[c]{3in}{
\epsfysize=2.0in\
\hspace{0.5in}\
\epsfbox{#4}
\vspace*{-3mm}\caption[]{#5}\label{#6} \
}\end{minipage}
\end{center}
\vspace*{-0.4in}\
\end{figure*}
}
}

\def\cfiguredouble[#1,#2,#3,#4]{
\begin{figure}
\vspace*{0.2in}\
\begin{center}
\begin{minipage}[c]{1.5in}{
\epsfxsize=1.5in\
\epsfbox{#1}
}\end{minipage}\hspace*{0.1in}\
\begin{minipage}[c]{1.5in}{
\epsfxsize=1.5in\
\vspace{0.1in}\epsfbox{#2}
}\end{minipage}\vspace*{-0.10in} \caption[]{#3}\label{#4}
\end{center}
\vspace*{-0.4in}\
\end{figure}
}

\def\wpfigure[#1,#2,#3,#4]{
\begin{figure*}
\vspace*{4mm}
\begin{center}

\includegraphics[width=#4]{#1} 

\vspace*{-3mm}\caption[]{#2
} \label{#3}

\vspace*{-5mm}
\end{center}
\end{figure*}}

\def\wprfigure[#1,#2,#3,#4,#5]{
\begin{figure*}
\vspace*{4mm}
\begin{center}

\includegraphics[width=#4, angle=#5]{#1} 

\vspace*{-3mm}\caption[]{#2
} \label{#3}

\vspace*{-5mm}
\end{center}
\end{figure*}}

\def\DoubleFigureWSlide[#1,#2,#3,#4,#5,#6,#7,#8,#9]{
\begin{figure*}
\vspace*{#9}
\begin{center}
\begin{minipage}{#4}
\includegraphics[width=#4]{#1}
\vspace*{-3mm}\caption{#2
}\label{#3}
\end{minipage}
\hspace{2em}
\begin{minipage}{#8}
\includegraphics[width=#8]{#5}
\vspace*{-3mm}\caption{#6
}\label{#7}
\end{minipage}
\vspace*{-5mm}
\end{center}
\end{figure*}
}

\def\DoubleFigureW[#1,#2,#3,#4,#5,#6,#7,#8]{
\begin{figure*}
\vspace*{0in}
\begin{center}
\begin{minipage}{#4}
\includegraphics[width=#4]{#1}
\vspace*{-3mm}\caption{#2
}\label{#3}
\end{minipage}
\hspace{2em}
\begin{minipage}{#8}
\includegraphics[width=#8]{#5}
\vspace*{-3mm}\caption{#6
}\label{#7}
\end{minipage}
\vspace*{-5mm}
\end{center}
\end{figure*}
}

\def\DoubleFigureWHack[#1,#2,#3,#4,#5,#6,#7,#8]{
\begin{figure*}
\vspace*{0in}
\begin{center}
\begin{minipage}{3in}
\includegraphics[width=#4]{#1}
\vspace*{-3mm}\caption{#2
}\label{#3}
\end{minipage}
\hspace{2em}
\begin{minipage}{3in}
\includegraphics[width=#8]{#5}
\vspace*{-3mm}\caption{#6
}\label{#7}
\end{minipage}
\vspace*{-5mm}
\end{center}
\end{figure*}
}

\def\ddcfigure[#1,#2,#3,#4]{
\begin{figure*}
\begin{center}
\begin{minipage}[c]{\columnwidth}{
\includegraphics[width=\columnwidth]{#1} 
}\end{minipage}\hspace{0.5in}\
\begin{minipage}[c]{\columnwidth}{
\includegraphics[width=\columnwidth]{#2} 
}\end{minipage} \caption[]{#3}\label{#4}
\end{center}
\vspace{1pt}
\end{figure*}
}

\def\ddcfigureSlide[#1,#2,#3,#4,#5]{
\begin{figure*}
\vspace*{#5}\
\begin{center}
\begin{minipage}[c]{3in}{
\includegraphics[height=3in]{#1} 
}\end{minipage}\hspace{0.5in}\
\begin{minipage}[c]{3in}{
\includegraphics[height=3in]{#2} 
}\end{minipage}\vspace*{-0.10in} \caption[]{#3}\label{#4}
\end{center}
\vspace*{-0.4in}\
\end{figure*}
}

\def\cxfigure[#1,#2,#3]{
\begin{figure}
\vspace*{4mm}
\begin{center}
 
\epsfxsize=2.5in\
\epsfbox{#1}\
 
\vspace*{-0.10in}\caption[]{#2
} \label{#3}
 
\vspace*{-5mm}
\end{center}
\vspace*{-2mm}
\end{figure}}

\newif\ifremark
\long\def\remark#1{
        \begingroup%
        \dimen0=\columnwidth
        \advance\dimen0 by -1in%
        \setbox0=\hbox{\parbox[b]{\dimen0}{\protect\em #1}}
        \dimen1=\ht0\advance\dimen1 by 2pt%
        \dimen2=\dp0\advance\dimen2 by 2pt%
        \vskip 0.25pt%
        \hbox to \columnwidth{%
                \vrule height\dimen1 width 3pt depth\dimen2%
                \hss\copy0\hss%
                \vrule height\dimen1 width 3pt depth\dimen2%
        }%
        \endgroup%
}

\usepackage{xcolor}

\definecolor{cyanish}{rgb}{0,0.8,1.0}
\definecolor{orange}{rgb}{1.0,0.5,0.0}
\definecolor{pink}{rgb}{1.0,0.47,0.6}
\definecolor{light-gray}{gray}{0.95}
\definecolor{jiancolor}{RGB}{0,153,153}
\definecolor{mygreen}{RGB}{50,200,50}
\definecolor{pink}{rgb}{1.0,0.47,0.6}

\definecolor{commentgreen}{rgb}{0.0,0.5,0.0}

\newcommand{\ignore}[1]{}

\newcommand{\reffig}[1]{Figure~\ref{#1}}
\newcommand{\refsec}[1]{Section~\ref{#1}}
\newcommand{\reftab}[1]{Table~\ref{#1}}
\newcommand{\reflns}[2]{Lines~\hyperref[#1]{\ref*{#1}-\ref*{#2}}}
\newcommand{\us}{\textmu{}s}
\newcommand{\x}[1]{$\times$}

\newcommand{\smahar}[1]{{{\textcolor{cyan}{{\small\sffamily $\blacktriangleright$ Suyash: #1 $\blacktriangleleft$}}}}}

\newif\ifcutforspace
\long\def\cutforspace#1{
\ifcutforspace%
        \begingroup%
        \dimen0=\columnwidth
        \advance\dimen0 by -1in%
        \setbox0=\hbox{\parbox[b]{\dimen0}{\protect{\em Cut For Space} #1}}
        \dimen1=\ht0\advance\dimen1 by 2pt%
        \dimen2=\dp0\advance\dimen2 by 2pt%
        \vskip 0.25pt%
        \hbox to \columnwidth{%
                \vrule height\dimen1 width 3pt depth\dimen2%
                \hss\copy0\hss%
                \vrule height\dimen1 width 3pt depth\dimen2%
        }%
        \endgroup%
\fi}

\newcommand{\nova}[1]{NOVA} 
\newcommand{\csym}[1]{\texttt{#1}}

\newcommand{\cfunc}[1]{\mbox{\csym{#1}\hspace{-0.1em}\csym{()}}}

\newcommand{\malloc}{\cfunc{malloc}}
\newcommand{\free}{\cfunc{free}}
\newcommand{\mmap}{\cfunc{mmap}}
\newcommand{\memcpy}{\cfunc{memcpy}}

\newcommand{\eg}{e.g.}

\newcommand{\uopsside}[3]{\multirow{#2}{*}{\sffamily\rotatebox[origin=c]{90}{\parbox[c]{#1cm}{\textbf{#3}}}}}

\newcommand{\xname}{RPCool}
\newcommand{\xdb}{CoolDB}

\newcommand{\libxname}{\texttt{librpcool}}
\newcommand{\mprotect}{\texttt{mprotect()}}
\newcommand{\load}{\texttt{load}}
\newcommand{\store}{\texttt{store}}
\newcommand{\seal}{\texttt{seal()}}
\newcommand{\release}{\texttt{release()}}
\newcommand{\tmpfs}{\texttt{tmpfs}}

\newcommand*\circledSolid[1]{\tikz[baseline=(char.base)]{
            \node[shape=circle,fill,inner sep=0.5pt] (char) {\textcolor{white}{#1}};}}

\newcommand{\tablefont}{\fontsize{8}{10}\selectfont}

\newcommand{\todo}[1]{{\color{red}TODO: #1}}

\newcommand{\NopLatRpcool}{1.5~\us{}}
\newcommand{\NopLatRpcoolRdma}{17.25~\us{}} %

\newcommand{\NopLatGrpc}{5.5~ms}
\newcommand{\NopLatErpc}{2.9~\us{}}
\newcommand{\NopLatZhangrpc}{10.9~\us{}}
 
\newcommand{\NopKrpsRpcool}{642.75}
\newcommand{\NopKrpsRpcoolRdma}{57.99}
\newcommand{\NopKrpsGrpc}{0.18}
\newcommand{\NopKrpsErpc}{334.03}
\newcommand{\NopKrpsZhangrpc}{99.69}

\newcommand{\createChannelbenchmark}{26.5~ms}
\newcommand{\destroyChannelbenchmark}{38.4~ms}
\newcommand{\connectBenchmark}{0.4~s}

\newcommand{\LatNopLatRpcoolkernelSecured}{2.6~\us{}}
\newcommand{\KrpsNopLatRpcoolkernelSecured}{377.79}

\newcommand{\cooldbsearchspeedup}{1.3}
\newcommand{\cooldbbuildspeedup}{4.7}

\usepackage{hyperref}

\definecolor{lightgray}{gray}{0.94}

\author{Suyash Mahar\textsuperscript{*}}
\affiliation{%
  \institution{UC San Diego}
  \city{San Diego}
  \state{CA}
  \country{USA}}

\author{Ehsan Hajyjasini\textsuperscript{*}}
\affiliation{%
  \institution{UC San Diego}
  \city{San Diego}
  \state{CA}
  \country{USA}}

\author{Seungjin Lee}
\affiliation{%
  \institution{UC San Diego}
  \city{San Diego}
  \state{CA}
  \country{USA}}

\author{Zifeng Zhang}
\affiliation{%
  \institution{UC San Diego}
  \city{San Diego}
  \state{CA}
  \country{USA}}

\author{Mingyao Shen}
\affiliation{%
  \institution{UC San Diego}
  \city{San Diego}
  \state{CA}
  \country{USA}}

\author{Steven Swanson}
\affiliation{%
  \institution{UC San Diego}
  \city{San Diego}
  \state{CA}
  \country{USA}}

\thanks{*These authors contributed equally to this work.}

\hypersetup{
    colorlinks = true,
    linkbordercolor = {white}
}

\begin{document}
\sloppy

\addtolength{\floatsep}{-1mm}
\addtolength{\textfloatsep}{-4mm}
\addtolength{\dbltextfloatsep}{-5mm}
\setlength{\abovecaptionskip}{0pt plus 1pt minus 2pt}
\setlength{\belowcaptionskip}{0pt plus 1pt minus 2pt}

\title{Telepathic Datacenters: Fast RPCs using Shared CXL Memory}

\date{}

\begin{abstract}
  Datacenter applications often rely on remote procedure calls (RPCs) for fast,
  efficient, and secure communication.  However, RPCs are slow, inefficient, and
  hard to use as they require expensive serialization and compression to
  communicate over a packetized serial network link. Compute Express Link 3.0
  (CXL) offers an alternative solution, allowing applications to share data
  using a cache-coherent, shared-memory interface across clusters of machines.

  \xname{} is a new framework that exploits CXL's shared memory
  capabilities. \xname{} avoids serialization by passing pointers to data
  structures in shared memory. While avoiding serialization is useful, directly
  sharing pointer-rich data eliminates the isolation that copying data over
  traditional networks provides, leaving the receiver vulnerable to invalid
  pointers and concurrent updates to shared data by the sender.  \xname{}
  restores this safety with careful and efficient management of memory
  permissions.  Another significant challenge with CXL shared memory
  capabilities is that they are unlikely to scale to an entire
  datacenter. \xname{} addresses this by falling back to RDMA-based
  communication.

  Overall, \xname{} reduces the round-trip latency by 1.93\x{} and 7.2\x{}
  compared to state-of-the-art RDMA and CXL-based RPC mechanisms,
  respectively. Moreover, \xname{} performs either comparably or better than
  other RPC mechanisms across a range of workloads.
\end{abstract}

\settopmatter{printfolios=true}
\maketitle
\pagestyle{plain}

\section{Introduction}

Communication within the datacenter needs to be fast, efficient, and secure
against unauthorized access, rogue actors, and buggy programs. Remote procedure
calls (RPCs)~\cite{grpc, thriftrpc} are a popular way of communicating between
independent applications and make up a significant portion of datacenter
communication, particularly among microservices. However, RPCs require
substantial resources to serve today's datacenter communication needs. For
instance, Google reports~\cite{google-rpc-study} that in the tail (\eg, P99),
requests spend over 25\% of their time in the RPC stack. One of the significant
sources of RPC latency is their need to serialize/deserialize and
compress/decompress data before/after transmission, which is especially
resource-intensive for pointer-rich data structures like trees and graphs.

Compute Express Link 3.0 (CXL)~\cite{cxl} promises to provide multi-host shared
memory, offering an exciting alternative by providing hardware cache coherency
among multiple compute nodes. Instead of serializing data structures and
transmitting them over the network, applications could share pointers to the
original data, significantly lowering their CPU usage.

However, sharing pointer-rich data in shared memory raises several safety
concerns.  Shared memory eliminates the traditional isolation of the sender from
the receiver that serialized networking provides. For example, the sender could
concurrently modify shared data structures while the receiver processes them,
leading to unsynchronized memory sharing between mutually distrustful
applications. This lack of synchronization can result in a range of potentially
dire consequences.

Another major challenge with CXL-based shared memory RPC is that CXL memory
coherence will likely be limited to rack-scale systems~\cite{cxl-switch} and
will almost certainly not span an entire datacenter. Thus, an RPC system that
works only at rack scale is just not suited for the datacenter. It must provide
a reasonable backup plan if CXL is not available.

\ignore{in a traditional datacenter
environment where two communicating applications are placed on different racks,
the application must dynamically switch RPC protocol, adding significant
development and programming overhead.}

Finally, using shared memory for communication results in challenges with
availability and memory management. For example, applications can leak shared
memory if they crash without relinquishing the memory.

To solve these issues, we propose \xname{}, a CXL shared memory-based RPC
library that exposes the benefits of shared-memory communication while
addressing the pitfalls described above. Using \xname{}, clients and servers can
directly exchange pointer-rich data structures residing in coherent shared
memory. \xname{} is the first RPC framework to implement a fast, efficient, and
scalable RPC framework while addressing the security and scalability concerns of
shared memory communication.

\xname{} provides the following features for improved RPC performance while
overcoming the limitations of CXL-based shared memory communication:

\begin{enumerate}[leftmargin=8mm]
\item\textit{Native pointer-rich data as RPC arguments.} \xname{} lets
applications send, receive, and share native pointer-rich data structures
without serialization.

  \item\textit{Preventing sender-receiver concurrent access.} \xname{} prevents
  the sender from modifying in-flight data by restricting its access to RPC
  arguments.

  \item\textit{Lightweight checks for invalid and wild pointers.}  \xname{}
  provides a lightweight sandbox to check for invalid or wild pointers while
  processing RPC arguments in shared memory.

  \item\textit{Seamless RDMA fallback.} \xname{} seamlessly switches to use RDMA
  to address CXL's scaling limitations while providing a unified RPC interface.

\item\textit{Shared memory management.} \xname{} can notify applications of
  shared memory failures, and limit shared memory consumption to prevent data
  loss or memory leaks.
\end{enumerate}

Using \xname{}, applications can construct pointer-rich data structures with a
\malloc{}/\free{}-like API and share them as RPC arguments. Clients can choose
whether to share the RPC arguments with other clients or keep them private to
the server and the client. To coordinate memory management and decide between
CXL and RDMA-based communication, \xname{} includes a global orchestrator. The
global orchestrator also manages connections and tracks shared memory regions
among applications.

We compare \xname{}'s performance against several other RPC frameworks,
including state-of-the-art RDMA, TCP, and CXL-based RPC frameworks. Overall,
\xname{} achieves the lowest round-trip time and highest throughput of any RPC
framework for no-op RPCs. To showcase \xname{}'s ability to share complex data
structures, we implemented a JSON-like document store and compared \xname{}
against eRPC~\cite{erpc}, gRPC~\cite{grpc}, and
ZhangRPC~\cite{zhang2023partial}. Our results show a \cooldbbuildspeedup{}\x{}
speedup for building the database and a \cooldbsearchspeedup{}\x{} speedup for
search operations compared to the fastest RPC frameworks.

We also evaluated \xname{}'s performance against TCP and UNIX domain sockets
using modified versions of MongoDB and Memcached. Across the two databases,
\xname{} significantly outperforms TCP in most workloads.  In the DeathStarBench
social network microservices benchmark, \xname{} performs on par with Thrift
RPC, as the benchmark's performance is constrained by the need to update various
databases on the critical path.

  Rest of the paper is structured as follows: \refsec{sec:background} presents
  the overview of RPCs and their limitations.  Next,
  \refsec{sec:cxl-transport-layer} discusses how CXL can alleviate these
  limitations. Section~\ref{sec:overview} and~\ref{sec:system} %
  presents our CXL-based RPC framework, \xname{} and various system details,
  respectively. In \refsec{sec:results}, we evaluate \xname{}'s
  performance. Finally, we discuss works related to \xname{} in
  \refsec{sec:related} and conclude in \refsec{sec:conclude}.

\section{RPCs in Today's World}
\label{sec:background}

Modern RPC frameworks grew from the need to make function calls across process
and machine boundaries, making programming distributed systems
easier~\cite{birrell1984implementing}. These RPCs provide an illusion of
function calls while relying on a layer cake of underlying technologies that
results in lost performance. For example, RPC frameworks waste a significant
number of CPU cycles on serializing and deserializing RPC arguments to send them
over traditional networking interfaces.

CXL offers the opportunity to rethink the design of RPC mechanisms.  To better
understand the attendant challenges, let us examine the structure and
limitations of modern RPC systems. Then, we will follow with a discussion of how
CXL-based shared memory can alleviate these problems.

RPCs provide an interface similar to a local procedure call: the sender makes a
function call to a function exported by the framework. The RPC framework
serializes the arguments and sends them over the network to the receiver. At the
receiver, the RPC framework deserializes the arguments and calls the appropriate
function.

While RPC frameworks provide a familiar abstraction for invoking a remote
operation, the underlying technology used results in several limitations:

First, to enable communication over transports like TCP/IP, RPC frameworks
serialize and deserialize RPC arguments and return values. This adds significant
overhead to sending complex objects (e.g., the lists and maps that make up a
JSON-like object in memory).

Second, most RPC frameworks do not support sharing pointer-rich data structures
due to different address space layouts between the sender and the
receiver. Applications can circumvent this by using ``smart
pointers''~\cite{zhang2023partial} or ``swizzling'' pointers, but both of these
add additional overheads.

Third, the underlying communication layer limits today's RPC frameworks'
performance. For example, the two most common RPC frameworks, gRPC~\cite{grpc}
and ThriftRPC~\cite{thriftrpc} rely on HTTP and TCP, respectively. Some RPC
frameworks like eRPC~\cite{erpc} exploit the low latency and high throughput of
RDMA to achieve better performance but are still limited by the underlying RDMA
network.

\section{CXL: A New Transport Layer.}
\label{sec:cxl-transport-layer}

\begin{figure}
  
\begin{minipage}{.18\textwidth}
  \centering \includegraphics[width=\linewidth]{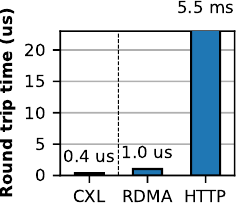}
  \captionof{figure}{RTT comparison of several communication protocols.}
  \label{fig:rpc-motivation}
\end{minipage}\hfill
\begin{minipage}{.28\textwidth}
  \centering
  \includegraphics[width=\linewidth]{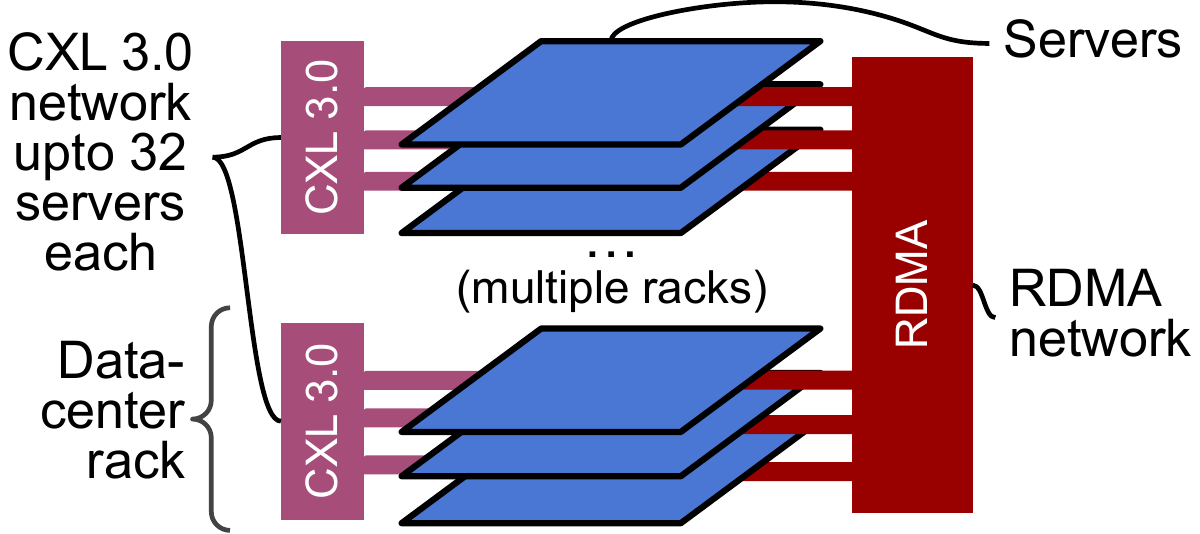}
  \captionof{figure}{Expected CXL v3+ in the datacenter alongside RDMA.}
  \label{fig:cxl-rpc-scenario-1}
\end{minipage}
\end{figure}

CXL 3.0 enables multiple hosts to communicate using fast, byte-addressable,
cache-coherent shared memory.  CXL-connected hosts will be able to map the same
region of shared memory in their address space~\cite{dax-cxl-lpc}, where updates
using \load{}/\store{} instructions from one host are visible to all other hosts
without explicit communication.

To understand how CXL might improve upon state-of-the-art RDMA-based RPC
frameworks, consider the round-trip latencies of CXL, RDMA, and HTTP protocols.
\reffig{fig:rpc-motivation} shows that based on the expected CXL access
latency~\cite{zhang2023partial}, a CXL-based RPC framework can potentially
improve the underlying communication layer's performance.

To better understand how an RPC framework can exploit CXL's features, we need to
first look into how CXL is expected to be deployed. In this work, we consider
the scenario where up to 32 servers with independent OSs are connected to a
single pool of shared memory using CXL, as shown in
\reffig{fig:cxl-rpc-scenario-1}.  Given the challenges of implementing
large-scale coherent memory, we assume that CXL memory sharing will not scale
beyond a single rack (\textasciitilde{}32--64 nodes).  We also expect CXL to
co-exist with conventional networking (\eg, TCP and RDMA).  Processes within a
rack can communicate over the CXL-based shared memory, avoiding expensive
network-based communication but can also communicate over RDMA to overcome CXL's
limited range.

This system architecture corresponds to a datacenter environment where
microservices are often spawned across multiple servers and communicate using
RPCs.

\section{\xname{}}
\label{sec:overview}

\ignore{\xname{} is a framework for fast and efficient RPC-based communication
  between CXL-connected hosts that can fall back to RDMA for compatibility and
  scalability. However, unlike traditional RPC mechanisms common today, \xname{}
  enables applications to share and access data without any serialization or
  copying while enabling the use of native pointers.}

\xname{} is a framework for fast and efficient RPC-based communication between
CXL-connected hosts. \xname{} enables applications to share and access data
without any serialization or copying, supports the use of native pointers, and
falls back to traditional networking to address the limited scalability of CXL.

To achieve this, \xname{} needs to address several challenges associated with
using shared memory for communication:

\begin{enumerate}[leftmargin=5mm]
\item \textit{Safely dereference native pointers.} \xname{} should enable
  applications to use native pointers without making them vulnerable to wild or
  invalid pointers.
\item \textit{Prevent concurrent access to shared data.} \xname{} should let
  applications take exclusive access to shared memory data to prevent malicious
  (or buggy) applications from concurrently modifying it.
\item \textit{Address the limited scalability of CXL.} \xname{} must enable
  applications to transparently use the same API to communicate beyond the
  limited scalability of CXL.
\item \textit{Shared memory coordination and failure handling.} \xname{}
  prevents distributed memory leaks and automatically reclaims memory after
  failures.
\end{enumerate}

\ignore{\xname{} overcomes these challenges by enabling applications to safely
dereference shared-memory native pointers, allows the receiver to request
exclusive access to shared data, scales beyond a single rack using RDMA to
overcome CXL's limitations, and coordinates memory management across
applications to prevent memory leaks and data loss.}

The following sections describe \xname{}'s architecture, its key components, and
how it achieves the above goals.

\subsection{\xname{} Architecture}

\begin{figure}
  \includegraphics[width=0.8\linewidth]{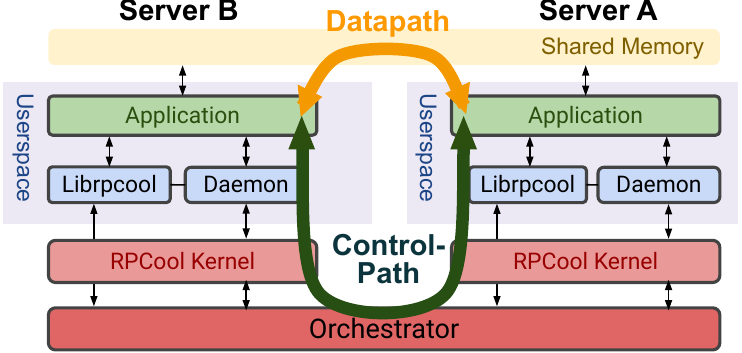}
  \caption{\xname{}'s System Architecture.}
  \label{fig:system-architecture}
\end{figure}

\xname{} uses CXL-based shared memory to safely communicate between processes
when possible and falls back to RDMA when necessary. The framework consists of
userspace components, \xname{}'s kernel, and a global orchestrator.

The userspace component includes a library (\libxname) and a trusted daemon,
which provide APIs for connecting to a specific server using \xname{}
``channels,'' sending/receiving RPCs, and managing shared memory objects. The
userspace components rely on \xname{}'s support in the kernel, which provides
\xname{}'s security guarantees and maps the shared memory regions into the
application's address space.

The global orchestrator tracks resources, supports POSIX-like access control
lists for the shared memory, and coordinates a globally unique address space for
the shared memory regions to enable the use of native pointers. The orchestrator
in \xname{} resembles an orchestrator commonly deployed for scaling and
restarting applications in a cluster or a datacenter. \ignore{In \xname{}, all
  resources managed by the orchestrator and the processes that can communicate
  with it comprise the \xname{} network.}

\begin{figure}
  \includegraphics[width=\linewidth]{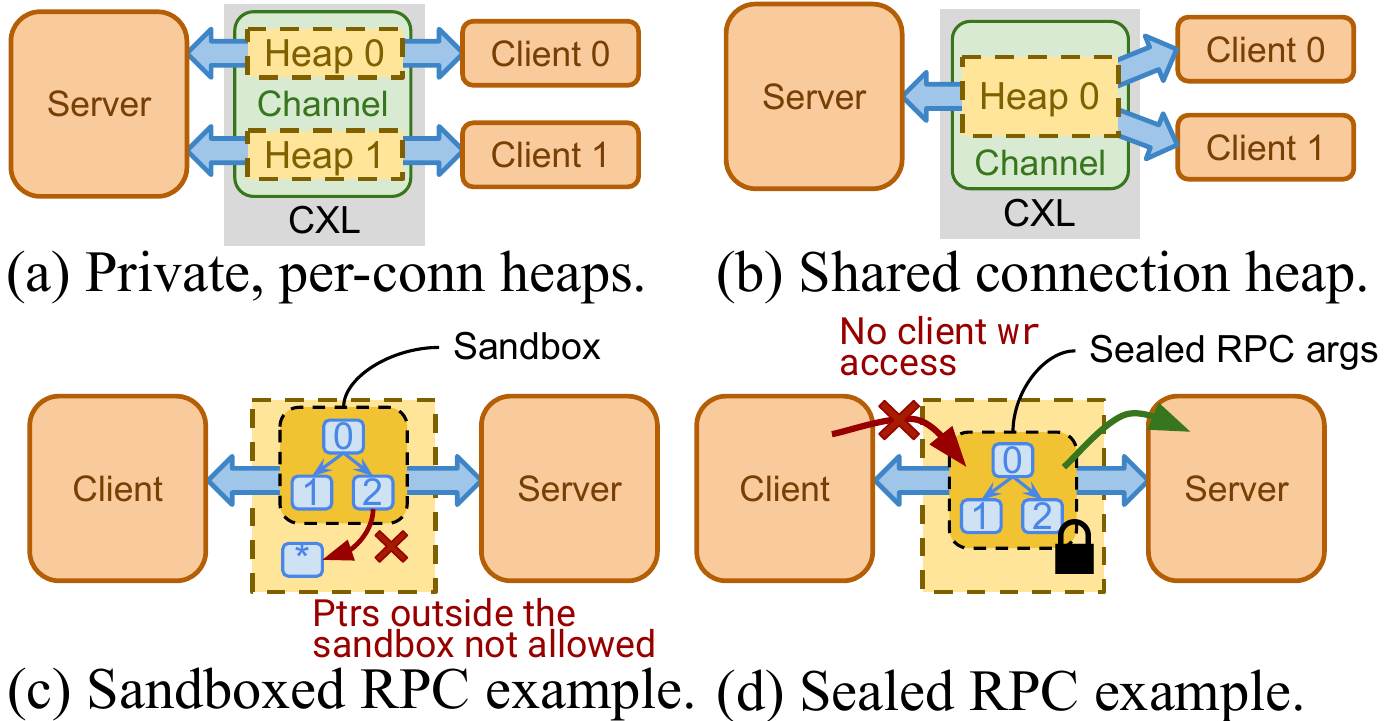}
  \caption{Channels, connections, and heaps in \xname{}.}
  \label{fig:channels-and-connections}
\end{figure}

\paragraph{Channels and Connections.} To allow applications to send RPCs, a server
creates a \emph{channel} that clients can connect to. Once connected, each
client receives a \emph{connection} object that provides access to the
connection's shared-memory heap. Channels in \xname{} automatically use either
CXL-based shared memory or fall back to RDMA, overcoming CXL's limited
scalability.

\paragraph{Shared memory heaps.} Each connection in \xname{} is associated with
a shared memory heap, enabling applications to allocate and share
objects. \reffig{fig:channels-and-connections}a--b shows how a single server can
serve multiple clients by using independent heaps that are private to each
connection (\reffig{fig:channels-and-connections}a) or by using a single shared
heap across multiple connections
(\reffig{fig:channels-and-connections}b). Connections start with a statically
sized heap and can allocate additional heaps if they need more space. When a
heap is created, the orchestrator assigns it a globally (in the cluster) unique
address where the heap will be mapped in a process's address space.  Giving each
heap a unique address space ensures that a client or server in cluster can
safely map it into its address space.

\paragraph{Seals and Sandboxes.}  
\xname{} includes support for sandboxes, which prevents any invalid or wild
pointers from causing invalid (or privacy-violating) memory access as the server
processes the RPC's arguments (\reffig{fig:channels-and-connections}c).

Moreover, \xname{} supports the ability to prevent the sender from concurrently
modifying RPC arguments as the receiver processes them. \xname{} achieves this
by dropping write access to the arguments for the sender, thus \emph{sealing}
the RPC (\reffig{fig:channels-and-connections}d). When an RPC is sealed, the
sender cannot modify the arguments until the receiver responds to the
RPC. Sandboxing and sealing are orthogonal and can be applied (or not) to
individual RPCs.

\paragraph{Shared memory management.} \xname{} provides a thread-safe memory
allocator to allocate/free objects from the shared memory heaps and several
STL-like containers such as \texttt{rpcool::vector}, \texttt{rpcool::string},
etc. These containers enable programmers to use the familiar STL interface for
allocating objects but do not preclude custom pointer-rich data structures,
e.g., trees or linked lists. The allocator and containers are based on
Boost.Interprocess~\cite{boost.interprocess}.

\xname{}'s orchestrator also requires each application that accesses shared
memory to periodically renew a lease so the orchestrator can track application
failures and clean up orphaned shared heaps.

Finally, to limit the amount of shared memory a process can amass, the
orchestrator enforces a system-administrator-defined shared-memory quota. The
quota limits the amount of heaps a process has access to at any time and
requires processes to return unused heaps to the orchestrator.

\subsection{Channels and Connections}
Channels and connections are the basic units for establishing communication
between two processes in \xname{}. Creating a channel in \xname{} is akin to
opening a port in traditional TCP-based communication. Once created, clients
``connect'' to the channel and get a connection object in return, enabling it to
send RPCs.  Every channel in \xname{} is identified by a unique, hierarchical
name and is registered with the orchestrator.

To enable participating processes to allocate and share objects, each connection
is associated with a region of the shared memory. Clients can choose to make
these heaps to be either private to a connection, or shared channel-wide.

\subsection{Shared Memory Safety Issues}
As discussed above, applications using \xname{} to share data over CXL-based
shared memory encounter two major safety issues. The first is the risk of wild
and invalid pointers. When processing an RPC, the receiver might dereference
such pointers, which could point to an invalid memory location and crash the
application, or alternatively, they could point to the receiver's private
memory, potentially leaking sensitive information. For example, a malicious
sender could exploit this by creating a linked list with its tail node pointing
to a secret key within the server, thereby extracting the key from a server that
computes some aggregate information about the elements in the list.

The second issue is concurrent access to shared data. When using shared memory
to share data structures, there is a risk that a sender might concurrently
modify an RPC's arguments while the server is processing them. In an untrusted
environment, a malicious sender could exploit this to extract sensitive
information from the receiver or crash it. While servers usually validate
received data, they must also ensure that the client cannot modify the shared
data once it has been validated.

\subsection{Preventing Unsafe Pointer Accesses Using Sandboxes}
\label{subsec:sandboxes}
\xname{} implements a lightweight sandbox to restrict received pointers from
pointing to any data outside the shared region while enabling applications to
use native pointers.

When processing a sandboxed RPC, a process enters the sandbox, losing access to
its private memory, and having access to only its shared memory heap and a set
of programmer-specified variables. If the process tries to access memory outside
the sandbox, it receives a signal that the process handles and uses to respond
to the RPC.

To minimize the cost of sandboxing incoming RPCs, \xname{} relies on Intel's
Memory Protection Keys (MPK)~\cite{sung2020intra}, avoiding the expensive
\mprotect{} system calls.  \refsec{subsec:sandboxes-impl} explains the details
of how \xname{}'s sandboxes work.

While we considered using non-standard pointers that enable runtime bound
checks, such pointers would limit compatibility with legacy software, compilers,
and debuggers and would have significant performance
overheads~\cite{mahar2024puddles}.

\subsection{Sealing RPC Data to Prevent Concurrent Accesses}
\label{subsec:seals}
In a trusted environment, the receiver can assume that the sender will not
concurrently modify the shared arguments while processing the RPC. However, in
scenarios where the receiver does not trust the sender, \xname{} must ensure
that the senders cannot modify an RPC's arguments while it is in flight.  There
are two attractive options to do this: First, the application can copy RPC
arguments, which works well for small objects, but for large and complex
objects, it is expensive. For these cases, \xname{} provides a faster
alternative---sealing the RPC arguments.  Seals in \xname{} apply to the
arguments of an in-flight RPC and prevent the sender from modifying them. The
sender uses the new \seal{} system call to seal the RPC and relinquish write
access to the RPC arguments when required by the receiver.  \libxname{} on the
receiver can then verify that the region is sealed by communicating with the
sender's kernel over shared memory. If not, \libxname{} would return the RPC
with an error.

When the receiver has processed the RPC, it marks the RPC as complete. The
sender then calls the \texttt{release()} system call, and its kernel verifies
that the RPC is complete before releasing the seal.

However, crisply defining which memory needs to be sealed requires special
attention in \xname{}'s design.

\xname{}'s solution is to provide scopes. Scopes provide a boundary around an
RPC's arguments, enabling applications to seal only the data needed for the
RPC. The alternative solution of sealing the entire heap prevents the sender
from having multiple in-flight RPCs, and sealing selective pages can result in
``false-sealing,'' where unrelated objects sharing a page are unnecessarily
sealed together.

Scopes in \xname{} are contiguous sets of pages that hold self-contained data
structures. Applications construct objects in scopes by allocating data directly
in the scope or copying them from the connection's heap. The sender can thus
send an RPC with arguments limited to a scope, sealing only the data needed for
the RPC. While scopes improve performance by limiting the pages sealed,
applications can still seal their entire heap, a tradeoff between the
performance and programming effort of managing and allocating data within
scopes.

\ignore{\paragraph{Combining Seals and Sandboxes}
\smahar{Proofread and improve the flow.}  In \xname{}, seals guarantee that the
sender no longer has access to a region of the connection's heap, however, the
receiver still has to ensure that the arguments received are contained within
the sealed region. This is because a malicious sender might send a pointer that
points to an unsealed region and could concurrently modify the memory location
alongside the receiver despite sending a sealed RPC.

A possible solution could be to check the pointers received by the receiver to
ensure that they are within the sealed region, however, locating and checking
every pointer is expensive and can result in significant overhead.

To solve this challenge, the receiver can use \xname{}'s sandbox along with
sealed RPCs, ensuring that a malicious sender does not send any pointer beyond
the sealed region.} 

\ignore{ When a receiver receives an RPC from an untrusted sender, it first
  verifies that the arguments are sealed and then creates a sandbox for the
  sealed region. This ensures that the sealed region is self-contained; that is,
  no pointer points to anything outside the region.}

\subsection{Handling Failures in \xname{}}

\begin{figure}
  \includegraphics[width=\linewidth]{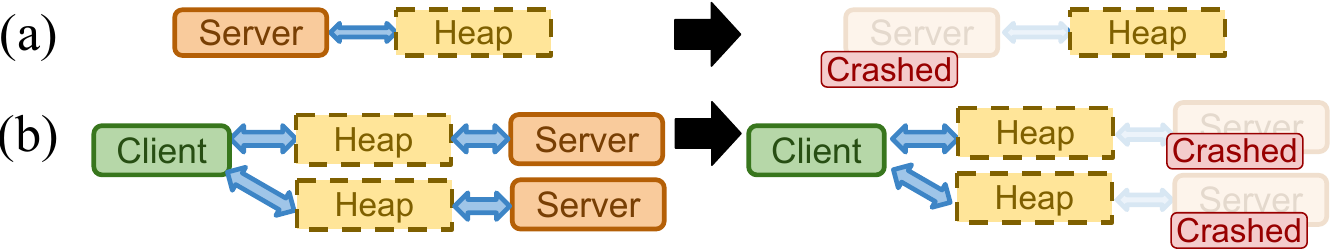}
  \caption{\textbf{Two possible failure scenarios in \xname{}.} (a) Server crash
    results in an orphaned heap. (b) Client left with heaps after multiple
    servers crash.}
  \label{fig:rpcool-failures}
\end{figure}

\xname{} must be able to deal with the two major shared-memory failure
scenarios: (a) orphaned heaps resulting from the crash of all applications
accessing a heap and (b) clients retaining heaps from failed connections,
leading to continued consumption of shared memory.

For example, if a server process that is not talking to any client dies, the
heaps associated with it are leaked, as no process manages it anymore
(\reffig{fig:rpcool-failures}a). Furthermore, consider a client application that
connects to multiple servers; if one of these servers fails, the client might
not free the associated heaps and retain a significant amount of shared memory
(\reffig{fig:rpcool-failures}b), consuming shared resources.

To address these challenges, \xname{} uses leases and quotas. Every time a
process maps a heap as part of a connection, it receives a lease from the
orchestrator. Applications using shared memory heaps periodically renew their
leases. When a process fails, the lease expires, and the orchestrator can notify
other participants and clean up any orphaned heaps. Upon a failure notification,
an application can either continue using the heap to access previously allocated
objects or release it if it is no longer needed, freeing up resources.

\subsection{RDMA Fallback}
While CXL enables applications to use multi-host shared memory to communicate,
it is unlikely to scale to large clusters~\cite{cxl-switch}. For deployment in
large clusters, \xname{} supports falling back to RDMA for communication between
hosts that cannot share memory via CXL.

When CXL is not an option, \xname{} replaces CXL's coherence mechanism with an
optimized RDMA-based software coherence system. \xname{} implements a minimalist
two-node RDMA-based shared memory, avoiding the expensive synchronization of
multi-node distributed shared memory (DSM) implementations like
ArgoDSM~\cite{argodsm}.

Whenever a node writes to a page, it gets exclusive access to the page by
unmapping it from all other nodes that have access to it. After the node has
updated the page, it can send an RPC to the other compute node, which can then
access the page at which \xname{} moves the page to the receiver.

\subsection{Example \xname{} Program}
\begin{figure}
  \includegraphics[width=\linewidth]{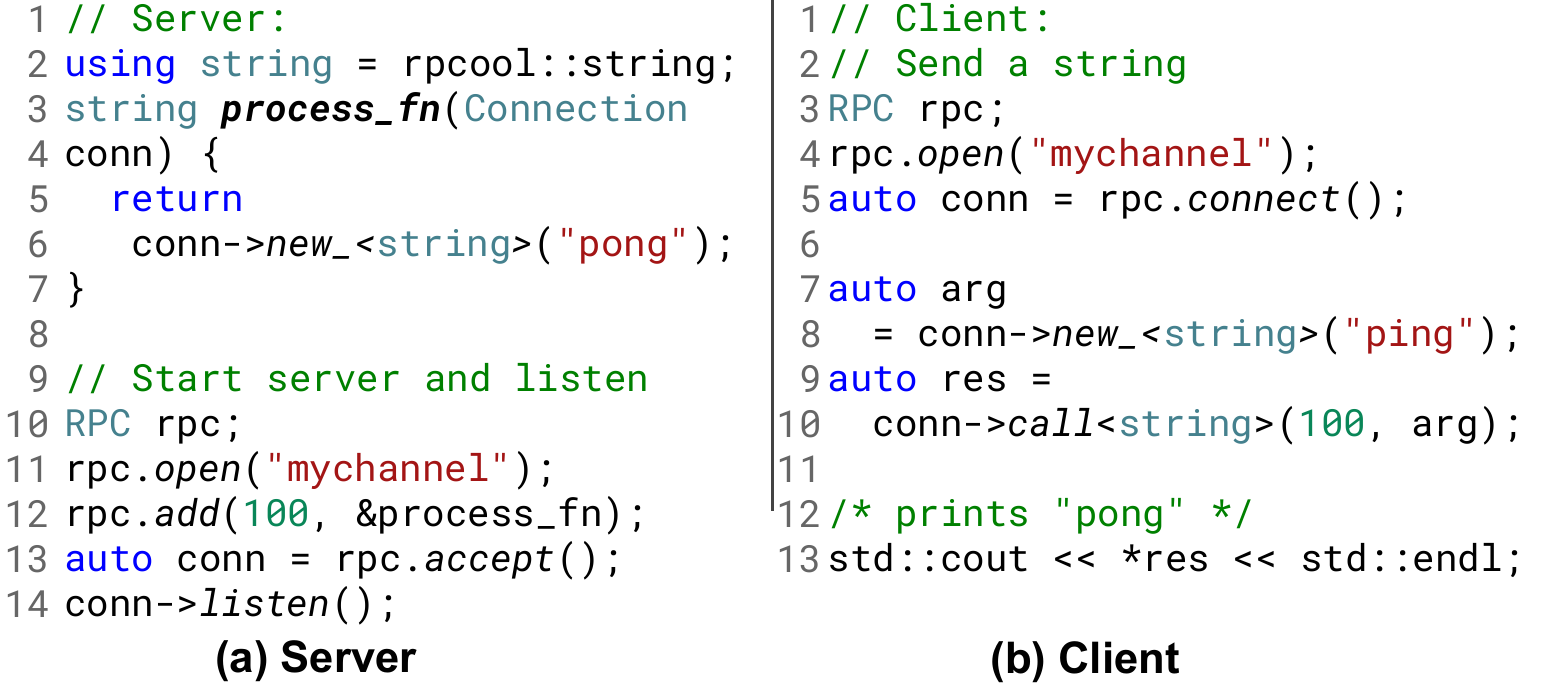}
  \caption{A simple ping-pong server using \xname{}.}
  \label{fig:code-example}
\end{figure}

\reffig{fig:code-example} shows the source code for an \xname{}-based server and
client that communicates over an \xname{} channel, \texttt{mychannel}. The
server registers \texttt{process\_fn()} (Line 12) that responds to the client's
ping requests. Once the function is registered, the server listens for any
incoming connections (Line 14).

Similarly, once the client has connected to the server (Line 5), it constructs a
new string in the connection's heap and calls the ping function on the server
(Line 9--10). Once the server responds to the request, the client prints the
result (Line 13).

\section{System Details}
\label{sec:system}

In this section, we look into \xname{}'s implementation details including
implementing low-overhead sandboxes and addressing the performance overhead of
sealing and RDMA fallback.

\subsection{Scopes}
\label{subsec:scope-impl}
\xname{} lets applications seal portions of a connection's heap by marking the
corresponding pages in the sender's address space as unwritable.  Changes to
memory permissions occur at page granularity, so disabling access to an RPC
argument might inadvertently disable access to other, unrelated, objects.  To
avoid this, we use scopes which ensure that these pages contain only the data
related to the RPC.  A scope is a dedicated range of contiguous pages within the
connection's heap.  Applications can allocate new objects in the scope using the
scope's memory management API or by copying in existing object data.

To create a scope, the programmer requests a scope of the desired size from the
connection's heap using the \texttt{Connection::create\_scope(size)}
API. \xname{} allocates the requested amount of memory from the connection's
heap and initializes the scope's memory allocator. The programmer can then
allocate or free objects within the scope's boundary.

An application can destroy scopes to free the associated memory or reset it to
reuse the scope. Once destroyed or reset, all objects allocated within the scope
are lost.

\subsection{Sandboxes}
\label{subsec:sandboxes-impl}

\begin{figure}
  \includegraphics[width=\linewidth]{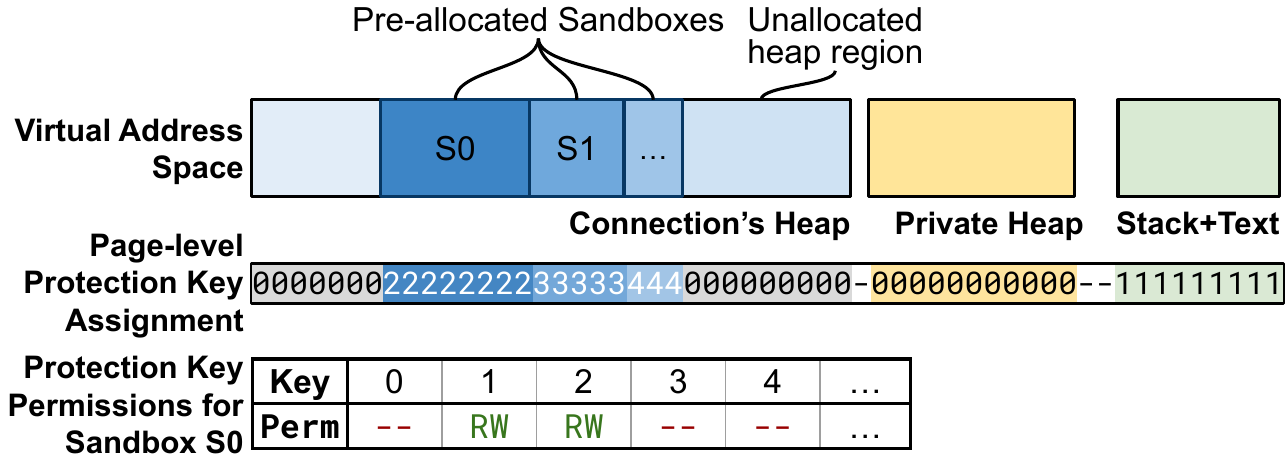}
  \caption{Preallocated sandboxes, their key assignment, and key permissions in
    \xname{}.}
  \label{fig:sandbox-working}
\end{figure}

\xname{} enables applications to sandbox an RPC by restricting the processing
thread's access to any memory outside of an RPC's arguments. This prevents the
applications from accidentally dereferencing pointers to private memory. To be
useful, \xname{}'s sandboxes must have low performance overhead, should allow
dynamic memory allocations despite restricting access to the process's private
memory, and permit selective access to private variables.

\paragraph{Low Overhead Sandboxes Using Intel MPK}
\xname{} uses Intel's Memory Protection Keys (MPK)~\cite{libmpk} to restrict
access to an application's private memory when in a sandbox, avoiding the much
more expensive \texttt{mprotect()} system call.  To use MPK, a process assigns
protection keys to its pages and then sets permissions using the per-cpu
\texttt{PKRU} register. In MPK, keys are assigned to pages at the process-level,
while permissions are set at the thread level.  Since MPK permissions are
per-thread, they enable support for multiple in-flight RPCs
simultaneously. Current Intel processors have 16 keys available.

Once a thread enters a sandbox, it uses Intel MPK to drop access to the
process's private memory and any part of the connection's heap except for the
sandboxed region. The receiver starts and ends sandboxed execution using the
\texttt{SB\_BEGIN(start\_addr, size\_bytes)} and \texttt{SB\_END} APIs.  The
receiver starts the sandbox with the same address and size as the scope used for
the RPC. However, \xname{} also supports sandboxing an arbitrary range of pages
within the connection's heap as required by an RPC.

To use Intel's MPK-based permission control, \xname{} assigns a key to each
region that needs independent access control, as shown in
\reffig{fig:sandbox-working}. \xname{} uses one key each for the application's
private memory, unsandboxed shared memory regions, and every sandbox.  Once a
key is assigned to a set of pages, \xname{} updates the per-thread \texttt{PKRU}
register entry to update their permissions.

When an application enters a sandbox, \xname{} drops access for all keys except
for the one assigned to the sandbox. If the sandboxed thread accesses any memory
outside the sandbox, the kernel generates a \texttt{SIGSEGV} that the process
can choose to propagate to the sender as an error.

\paragraph{Dynamic Allocations in Sandboxes}
As the sandboxed thread no longer has access to the process's private memory,
the thread cannot allocate objects in it.  However, the application may need to
allocate memory from \texttt{libc} using \malloc{}/\free{} or invoke a library
from within the sandbox that allocates private memory internally.

To address this, \xname{} redirects sandboxed \texttt{libc} \malloc{}/\free{}
calls to a temporary heap instead of the process's private heap. After the
sandbox exits, data in this temporary heap is lost. However, redirecting memory
allocations works only for libraries and other APIs that free their memory
before returning and do not maintain any state across calls. To safely use
stateful APIs over pointer-rich data, an application can validate the pointers
in a sandbox before calling the stateful API outside the sandbox.

\paragraph{Accessing Data Outside the Sandbox}
When in a sandbox, an application cannot access the connection's private heap,
however, in some cases applications might require access to certain private
variables to avoid entering and exiting the sandbox multiple times to service an
RPC call. To address this, \xname{} supports copying programmer-specified
private variables into the sandbox's temporary heap.  To copy a private
variable, the programmer specifies a list of variables in addition to the region
to sandbox when starting a sandbox: \texttt{SB\_BEGIN(}\texttt{region,}
\texttt{var0,} \texttt{var1...)}.

\paragraph{Optimizing Sandboxes}
Although changing permissions using Intel MPK takes tens of nanoseconds,
assigning keys to pages has similar overheads as the \texttt{mprotect()} system
call~\cite{libmpk}. To avoid assigning keys to on-demand sandboxes, \xname{}
reserves up to 14 pre-allocated or \emph{cached} sandboxes of varying sizes with
pre-assigned keys. This is limited by the number of protection keys
available. \xname{} reserves 2 keys for the private heap and unsandboxed
regions, respectively. \ignore{As applications are expected to use scopes for
  RPCs in an untrusted environment, the receiver can require the sender to use
  one of the pre-allocated sandbox regions as the scope for RPCs. This ensures
  that the receiver can use the preallocated sandbox, avoiding expensive system
  calls to assign protection keys to sandbox an RPC's arguments.}  To service a
request for an uncached sandbox region, \xname{} waits for an existing sandbox
to end, if needed, and reuses its key. This enables \xname{} to dynamically
create sandboxes without being limited to 14 pre-allocated sandboxes, albeit at
the cost of reassigning protection keys.

\subsection{Sealing Heaps}
\ignore{As \xname{} uses shared memory to share data structures, both the sender
  and the receiver have concurrent access to the shared data.  This leaves the
  receiver vulnerable as the sender can modify the shared data while the
  receiver is reading/writing to it.}

\ignore{In a naive, single-machine RPC scenario, \xname{} can address this by enabling
the receiver to request the OS to unmap the corresponding pages from the
sender's address space until the RPC is processed. However, as RPCs are expected
to run across operation system boundaries, preventing concurrent access by
multiple applications requires coordination among multiple operation systems.

To solve this problem, \xname{} enables support for sealing the arguments of an
RPC and preventing the sender from concurrently modifying them while an RPC is
in flight.}

\xname{}'s seal implementation should prevent the sender from concurrently
modifying an RPC's argument and should enable the receiver to verify the seal
before processing an RPC. This section describes how \xname{} implements its
sealing mechanism to achieve these features with high performance.

\paragraph{Seal Implementation.}  \xname{} enables the sender to enable sealing
on a per-request basis and specify the memory region associated with the
request.  When a sender requests \xname{} to seal an RPC, \libxname{} calls a
new \seal{} system call. In response, the kernel makes the corresponding pages
read-only for the sender and writes a seal descriptor to a sender-read-only
region in the shared memory. The receiver proceeds after it checks whether the
region is sealed by reading the descriptor.

Once an RPC is processed, the sender calls the \release{} system call and the
kernel checks to ensure the RPC is complete and releases the seal. The
descriptors are implemented as a circular buffer, mapped as read-only for the
sender but with read-write access for the receiver. These asymmetric permissions
allow only the receiver to mark the descriptor as complete and the sender's
kernel to verify that the RPC is completed before releasing the seal.

Further, as an application can have several seal descriptors active at a
given point in time, the sender also includes an index into the descriptor
buffer along with RPC's arguments.

\begin{figure}
  \includegraphics[width=\linewidth]{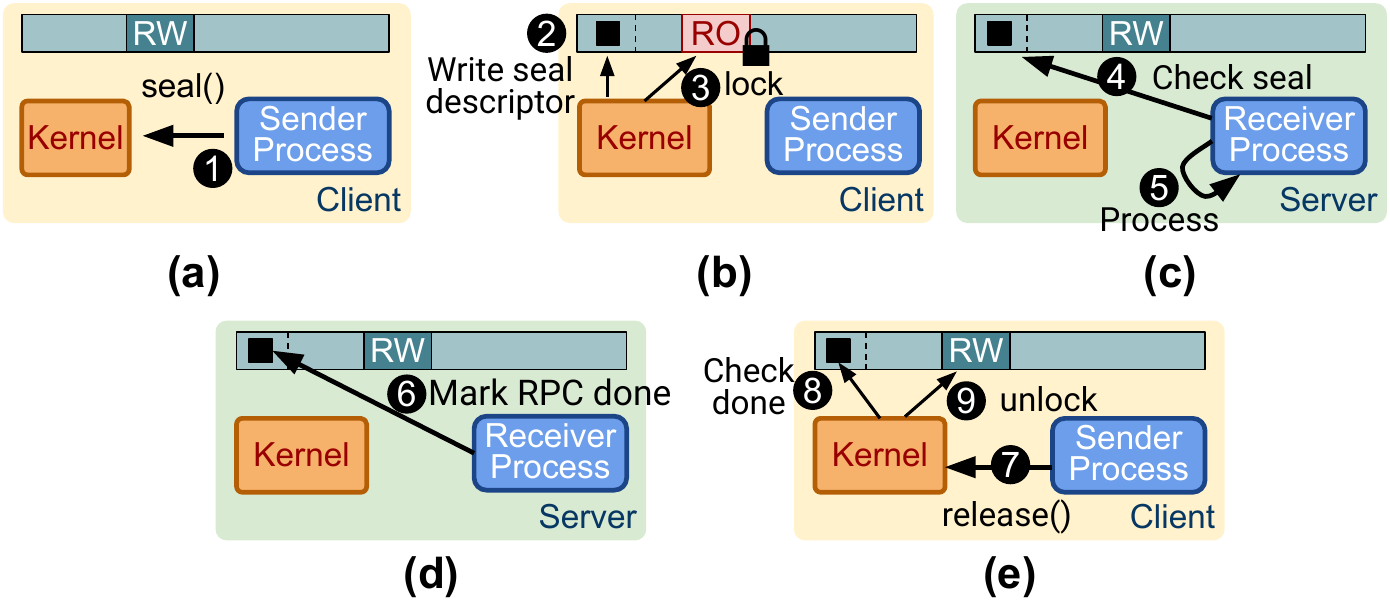}
  \caption{\textbf{Sealing mechanism overview}. The client sends a sealed RPC,
    and the receiver process checks the seal and processes it. Once processed,
    the receiver marks the RPC as completed, and the sender releases the seal.}
  \label{fig:sealing-mechanism-overview}
\end{figure}

\paragraph{Example.} \reffig{fig:sealing-mechanism-overview} describes the
sealing mechanism. Before sending the RPC, the sender calls the \texttt{seal()}
system call \circledSolid{1} with the region of the memory to seal. Next, the
sender's kernel writes the seal descriptor \circledSolid{2}, followed by
locking the corresponding range of pages by marking them as read-only in the
sender's address space \circledSolid{3}.

Once sealed, the RPC is sent to the receiver.  If the receiver is expecting a
sealed RPC, it uses \texttt{rpc\_call::isSealed()} to read and verify the seal
descriptor \circledSolid{4}, and processes the RPC if the seal is valid. After
processing the request \circledSolid{5}, the receiver marks the RPC as complete
in the descriptor \circledSolid{6} and returns the call. Next, when the sender
receives the response, it asks its kernel to release the seal
\circledSolid{7}. The kernel verifies that the RPC is complete \circledSolid{8}
and releases the region by changing the permissions to read-write for the range
of pages associated with the RPC \circledSolid{9}.

\paragraph{Optimizing Sealing}
Repeatedly invoking \seal{} and \release{} incurs significant performance
overhead as they manipulate the page table permission bits and evict TLB
entries~\cite{amit2020don}. To mitigate this, \xname{} supports scope pools that
batch \release{} calls for multiple scopes. Batching releases amortize the
overhead across an entire batch, resulting in fewer TLB shootdowns.

To use batched release, applications pop a scope from the pool, allocate RPC
arguments within this scope, and send a sealed RPC. Upon the RPC's returns, if
the application does not immediately need to modify the RPC arguments, it can
opt to release the seal in a batch. Batched releases work best when the
application does not need to modify the sealed arguments until the batch is
processed. However, if needed, the application can invoke \release{} and release
the seal on the scope. In \xname{}, each application independently configures
the batch release threshold, with a threshold of 1024 achieving a good balance
between performance and resource consumption.

\subsection{Leases and Quotas}
\label{subsec:leases-and-quotas}
Applications that allocate shared memory and use it to communicate must
coordinate among themselves to prevent orphaned heaps and should be notified of
server failures. There are three scenarios that require careful memory
management and failure coordination among processes participating in shared
memory communication.

The first is process failure notifications. When any of the communicating
processes fail, other processes should be notified of the failure. This
notification ensures that clients can perform appropriate housekeeping measures
to clean up any partial states associated with a failed server.

Second, in the case of a total failure where multiple processes crash, but the
memory node is alive, the system must reclaim memory to prevent memory
leaks. Third, \xname{} needs to handle scenarios where if one or more servers
that a client is communicating with crash, the client could continue using the
associated heaps, resulting in the client potentially using up a large portion
or all of the shared memory.

\paragraph{Leases}
\xname{} notifies applications if the server they are communicating with fails
and garbage collects orphaned heaps. \xname{} achieves this by requiring a lease
every time an application maps a connection's heap. Orchestrator uses these
leases to track which processes have failed and can notify other applications
sharing the memory regions. \xname{} creates a lease for each heap, and
\libxname{} periodically and automatically renews the lease while the
application is running and using the memory.

If the server for a channel fails, the lease expires and the orchestrator
notifies all clients connected to the channel of the failure. The clients can
continue to access the heap memory but can no longer use it for
communication. They can also close the channel. When the last process accessing
the heap closes the connection, the orchestrator reclaims the heap.

\paragraph{Quotas}
\xname{} supports shared memory quotas to limit applications from mapping a
large amount of shared memory into its address space. \xname{}'s orchestrator
enforces this configurable quota at the process level. A heap mapped into
multiple processes counts against all of their quotas. If mapping a new heap to
a process's address space would exceed its quota, the process would need to
close enough existing channels to map the new heap.

\subsection{The Daemon and The Kernel}
In \xname{}, each operating system runs a trusted daemon on start that is
responsible for handling all connection and channel-related requests, as well as
controlling access to them in coordination with the orchestrator.

The daemon is the only entity in \xname{} that makes system calls to map or
unmap a connection's heap into a process's address space. Consequently, every
application must communicate with the daemon to open and close connections or
channels.  Although applications are permitted to make \seal{} and \release{}
calls, they are not allowed to call \mprotect{} on the connection's heap
pages. This restriction prevents the application from bypassing kernel checks
for releasing sealed pages.

\ignore{An alternative would be to enforce access control directly in
  the kernel. However, such an implementation would require implementing
  communication between the kernel and the orchestrator, which would increase
  the potential attack surface for the kernel.}

\subsection{RDMA Fallback}
\xname{} includes support for automatic RDMA fallback for communication links
that span CXL-connected machine domains.  While applications could use
traditional RPC frameworks like ThriftRPC or gRPC to bridge the gap, this leads
to additional programming overhead as the programmer needs to pick the API
depending on where the target service is running.  Moreover, \xname{} cannot
transparently fall back to an existing RPC system because none of them support
sending pointer-based data structures.

\xname{} addresses these limitations by implementing a simple RDMA-based shared
memory mechanism that is optimized for RPCool's pattern of memory sharing. Where
either a server or a client has exclusive access to a shared memory page. When a
server attempts to access the data on a page using \load{}/\store{}
instructions, the instruction succeeds if the server has exclusive ownership of
the page. If not, the server triggers a page fault, fetches the page from the
client, and re-executes the instructions once mapped. Once fetched, the page is
marked as unavailable on the client, and it would need to request the page back
from the server in order to access the page.

\ignore{One possible solution to address \xname{}'s limited one-to-one communication
over RDMA would be to implement a full, multi-node distributed shared memory
(DSM), such as ArgoDSM~\cite{kaxiras2015turning}. Unfortunately, we found that
using a full DSM results in a significant performance impact, thus, we implement
a simpler two-node shared memory.}

\paragraph{Programming Interface.}
\xname{} over RDMA supports communication only between one server and one
client. Consequently, \xname{} also does not support simultaneous access to a
heap over both CXL and RDMA. While \xname{} over RDMA only supports two-node
communication, all other programmer-facing interfaces are identical to
\xname{}'s CXL implementation, e.g., allocating and accessing shared objects.

This limitation exists because when a process wants exclusive access to a page
shared over RDMA, \xname{} must unmap the corresponding page from all other
processes across the datacenter that have access to it, which adds significant
performance overheads and system complexity.

To address this limitation, \xname{} includes support for deep-copying
pointer-rich data structures between connection heaps using the
\texttt{conn.copy\_from(ptr)} API. \texttt{copy\_from()} automatically traverses
a linked data structure using Boost.PFR~\cite{boost.pfr} and deep copies to the
connection's heap, allowing applications to interoperate between connections of
different types without significant programming overhead.

\paragraph{Sealing and Sandboxing with RDMA Fallback}
Sealing and sandboxing for RDMA-based shared memory pages works similarly to
\xname{}'s CXL implementation.

When a sender sends a sealed RPC, the corresponding pages are marked as
read-only in its address space, preventing any modifications by the sender while
the RPC is in-flight. Further, to process an incoming RPC over RDMA fallback,
the application can create a sandbox over the RPC's arguments in the same manner
as it would for processing an RPC over CXL-based shared memory.

\subsection{Object and Heap Ownership}
\xname{} provides ownership guarantees that are similar to multiprocess shared
memory. Any application with access to a channel can allocate or free its
objects. When the last process with access to a channel heap closes it, the heap
is automatically freed. \xname{} does not restrict applications' ability to
manage object or heap ownerships. For instance, an application can utilize
programming language level constructs to manage object ownership and lifetime.

\subsection{Busy Waiting for RPCs}
\label{subsec:busy-wait}
\xname{} uses busy waiting to monitor new RPCs and their completion
notifications.  However, busy waiting can lead to excessive CPU utilization. To
mitigate this issue, \xname{} introduces a brief sleep period between busy wait
iterations. Specifically, \xname{} skips sleeping between iterations if
the CPU load is less than 25\%, sleeps for 5~\us{} if the CPU load is between
25\% and 50\%, and sleeps for 150~\us{} if the CPU load exceeds 50\%. We observe
that this achieves a good balance between CPU load and performance.


\widowpenalties 1 100

\section{Results}
\label{sec:results}
We evaluate \xname{} to understand and contrast its raw latency and throughput
with other RPC frameworks and how different \xname{} features affect its
performance.

To understand how \xname{} performs in real-world workloads, we integrate
\xname{} with several applications like Memcached~\cite{memcached},
MongoDB~\cite{mongodb}, DeathStarBench~\cite{deathstarbench}, and a new
document store, \xdb{}. For these experiments, we use several RPC mechanisms
ranging from TCP/IP-based RPC frameworks like Google's gRPC~\cite{grpc} and
Apache's ThriftRPC~\cite{thriftrpc}, RDMA-based state-of-the-art
eRPC~\cite{erpc}, and a failure-resiliency focused CXL-based RPC framework by
Zhang et al.~\cite{zhang2023partial}.  Across the experiments, \xname{} refers
to the CXL-only version, while \xname{} (RDMA) is \xname{} running over RDMA,
and \xname{} (Secure) is \xname{} over CXL with sealing and sandboxing turned
on.

\subsection{Evaluation Configuration}
\label{subsec:evaluation-configuration}

As CXL 3.0 devices are not commercially available, we use a dual-socket machine
to emulate CXL's access latency. \xname{} maps all connection heaps to the far
node, which has all its CPUs marked offline in the kernel. For RDMA, we use two
servers with direct-attached Mellanox CX-5 NICs. For the TCP experiments, we use
the NIC in ethernet mode, enabling TCP traffic over the RDMA NICs
(IPoIB~\cite{ipoib}). All CXL experiments use two Intel Xeon Gold 6230 with
192~GiB of DRAM while RDMA experiments use a single Intel Xeon Gold 6230 with
96~GiB of DRAM. Unless stated otherwise, all experiments are run on the v6.1.37
of the Linux kernel with adaptive sleep between busy-wait iterations
(\refsec{subsec:busy-wait}).

\subsection{Microbenchmarks}

This section compares the performance of RPCool's basic operations and the
overheads of its key mechanisms.

\begin{table*}
  \tablefont{}
  \centering
  \begin{subtable}{\textwidth}
    \centering
    \setlength{\tabcolsep}{0.5em}    
    \fontsize{8}{10}
    \selectfont
      \begin{tabular}{|l|l|l|l|l|l|l|}
    \hline
    \textbf{Framework}            & \xname{}         & \xname{} (Seal+Sandbox)          & \xname{} (RDMA)      & eRPC~\cite{erpc} & ZhangRPC~\cite{zhang2023partial} & gRPC~\cite{grpc} \\\hline\hline
    \textbf{No-op Latency}        & \NopLatRpcool{}  & \LatNopLatRpcoolkernelSecured{}  & \NopLatRpcoolRdma{}  & \NopLatErpc{}    & \NopLatZhangrpc{}                & \NopLatGrpc{}    \\\hline
    \textbf{Throughput (K req/s)} & \NopKrpsRpcool{} & \KrpsNopLatRpcoolkernelSecured{} & \NopKrpsRpcoolRdma{} & \NopKrpsErpc{}   & \NopKrpsZhangrpc{}               & \NopKrpsGrpc{}   \\\hline
    \textbf{Transport}            & CXL              & CXL                              & RDMA                 & RDMA             & CXL                              & TCP              \\\hline
  \end{tabular}

    \caption{Latency and throughput comparison among \xname{}, RDMA-based eRPC,
      failure-resilient CXL-based Zhang-RPC, and gRPC.}
    \label{tab:noop-latency}
  \end{subtable}

  \begin{subtable}{\textwidth}
    \centering
    \fontsize{8}{9}
    \selectfont
      \begin{tabular}{|p{0.6cm}|l|l|p{8cm}|}
    \hline
                                                  & \textbf{Operation}                          & \textbf{Mean Latency}           & \textbf{Description}                                                \\\hline
    \uopsside{1.5}{6}{\xname{} Ops}               & No-op \xname{} RPC (CXL)                    & \NopLatRpcool{}                 & RTT for \xname{} no-op RPC over CXL.                                \\
                                                  & No-op \xname{} RPC (RDMA)                   & \NopLatRpcoolRdma{}             & RTT for \xname{} no-op RPC over RDMA.                               \\
                                                  & No-op Sealed+Sandboxed RPC (CXL, 1 page)    & \LatNopLatRpcoolkernelSecured{} & RTT latency for \xname{} with seal and a cached sandbox over CXL.   \\\cline{2-4}
                                                  & Create Channel                              & \createChannelbenchmark{}       & Channel creation latency                                            \\
                                                  & Destroy Channel                             & \destroyChannelbenchmark{}      & Channel destruction latency                                         \\
                                                  & Connect Channel                             & \connectBenchmark{}             & Latency to connect to an existing channel                           \\\hline
    \uopsside{1}{4}{Sandbox Ops}                  & Cached Sandbox Enter+Exit (1 page)          & 0.35~\us{}                      & Enter and exit a single sandbox with a single shared memory page.   \\
                                                  & Cached Sandbox Enter+Exit (1024 page)       & 0.35~\us{}                      & Enter and exit a single sandbox with 1024 shared memory pages.      \\
                                                  & Cached Multiple Sandbox Enter+Exit (1 page) & 0.47~\us{}                      & Enter and exit 8 sandboxes, no protection key reassignment.         \\
                                                  & Uncached Sandbox Enter+Exit (1 page)        & 25.57~\us{}                     & Enter and exit 32 sandboxes, requires reassigning protection keys.  \\\hline
    \uopsside{1.5}{6}{Seal/Release, \& \memcpy{}} & Seal + standard release, no RPC (1 page)    & 1.1~\us{}                       & Seal and release a single shm page without sending an RPC.          \\
                                                  & Seal + standard release, no RPC (1024 page) & 3.46~\us{}                      & Seal and release 1024 shm pages without sending an RPC.             \\\cline{2-4}
                                                  & Seal + batch release, no RPC (1 page)       & 0.65~\us{}                      & Seal and release in batch a single shm page without sending an RPC  \\
                                                  & Seal + batch release, no RPC (1024 page)    & 2.95~\us{}                      & Seal and release in batch 1024 shm page without sending an RPC      \\\cline{2-4}
                                                  & Remote-remote \memcpy{} (1 page)            & 1.26~\us{}                      & \memcpy{} latency for remote node to remote node copy (1 page).     \\ 
                                                  & Remote-remote \memcpy{} (1024 page)         & 2308.23~\us{}                   & \memcpy{} latency for remote node to remote node copy (1024 pages). \\ \hline
  \end{tabular}

    \caption{Comparison of various \xname{} operations, repeated 2 million
      times.}
    \label{tab:microbench}
  \end{subtable}
  \caption{Microbenchmark performance and \xname{} operation.}
\end{table*}

\paragraph{No-op Round Trip Latency and Throughput} \reftab{tab:noop-latency}
compares \xname{}'s CXL, secured, and RDMA variants against several RPC
frameworks and shows that \xname{} significantly outperforms all other RPC
frameworks by a wide margin. Unlike \xname{}, Zhang RPC attaches an 8-byte
header to every CXL object and uses fat pointers (\texttt{CXLRef}) for
references. Thus, simple operations like constructing a tree data structure
require creating a CXL object and a \texttt{CXLRef} per tree node. Further,
assigning a node as a child requires the programmer to call a special
\texttt{link\_reference()} API, adding overhead on the critical path.

\paragraph{\xname{} Operation Latencies} Next, we look at the latency of
\xname{}'s features in \reftab{tab:microbench}. \xname{} in CXL mode takes only
\NopLatRpcool{}, while it takes \NopLatRpcoolRdma{} over RDMA. For CXL, this
latency increases to \LatNopLatRpcoolkernelSecured{} when sealing and sandboxing
a single page.

\xname{}'s cached sandboxes (i.e., sandboxes with pre-assigned protection key)
have very low enter+exit latency at 0.35~\us{}. This latency grows to
25.57~\us{} when the sandbox is not cached and \xname{} needs to reassign
protection keys and set up the sandbox's heap.

Finally, using \reftab{tab:microbench}, we look at the latency of \memcpy{} to
compare it against the cost of sealing+sandboxing, which includes sealing a
page, starting a sandbox over it, and finally releasing it. This is because
applications can copy RPC arguments to prevent concurrent accesses from the
sender without using sealing+sandboxing. We observe that for more than two
pages, sealing+sandboxing is faster than \memcpy{} (1.45~\us{} vs
1.5~\us{}). This suggests that for data smaller than two pages, applications
should use \memcpy{}, while for data larger than two pages, applications should
use sealing+sandboxing.

\subsection{Applications}
To understand how \xname{} performs integrated with real-world workloads, we
compare several applications' performance using \xname{} and other RPC
mechanisms. Overall, we observe that \xname{}'s low-latency RPCs result in
significant performance improvement over traditional RDMA- and TCP-based
networks.

\paragraph{Memcached}

\newcommand{\wrkldrot}{\parbox[t]{2mm}{\multirow{6}{*}{\rotatebox[origin=c]{90}{\textbf{Workload}}}}}

\begin{figure} %
  \includegraphics[width=\linewidth]{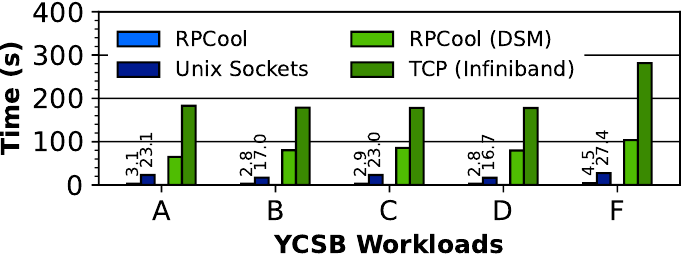}
  \caption{Memcached running the YCSB benchmark.}
  \label{fig:memcached-ycsb}
\end{figure} %
\begin{figure} %
  \includegraphics[width=\linewidth]{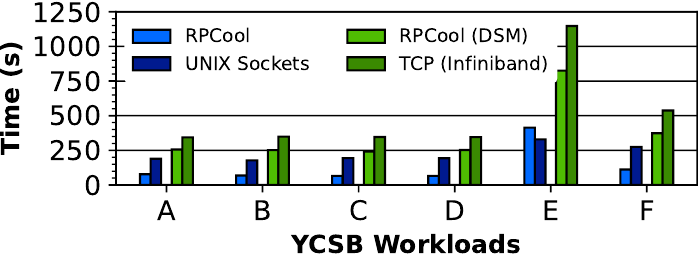}
  \caption{MongoDB running the YCSB benchmark.}
  \label{fig:mongodb-ycsb}
\end{figure}

\reffig{fig:memcached-ycsb} shows the execution time of memcached running the
YCSB benchmark~\cite{ycsb}. \xname{}'s CXL implementation significantly and
consistently outperforms UNIX domain sockets with a speedup of at least
6.0$\times$, while the DSM implementation outperforms TCP over Infiniband by at
least 2.1$\times$. As memcached transfers small amounts of non-pointer-rich
data, it uses \memcpy{} instead of sandboxing and sealing for isolation.

For each YCSB workload, we load Memcached with 100 thousand keys and run 1
million operations. Since Memcached is a key-value store, it does not support
SCAN operation and thus, it cannot run YCSB's E workload~\cite{ycsb-scan}.

\paragraph{MongoDB}
\reffig{fig:mongodb-ycsb} shows the execution time comparison of MongoDB using
\xname{} vs its built-in UNIX domain socket-based communication.  Across the
workloads, \xname{}'s CXL implementation outperforms UNIX domain sockets in all
workloads except the workloads E. Moreover, \xname{}'s DSM implementation
outperforms TCP over Inifiniband across all workloads by at least 1.34$\times$.

Like Memcached, we evaluate MongoDB with 100k keys and 1 million operations for
each workload and do not implement sealing+sandboxing as MongoDB internally
copies the non-pointer-rich data it receives from the client.

\begin{figure}
\includegraphics[width=\linewidth]{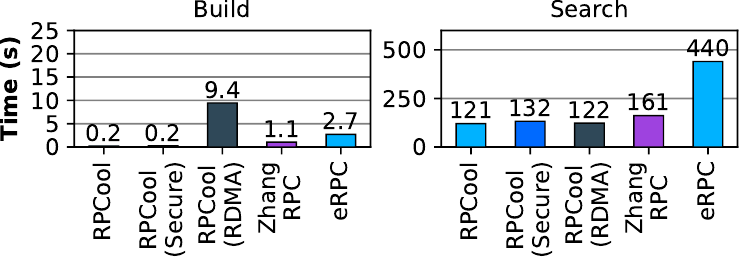}
\caption{\xdb{} execution time comparison for building the database
  (\textit{build}) and for searching keys (\textit{search}).}
  \label{fig:cooldb-perf}
\end{figure}

\paragraph{\xdb{}}
\xdb{} is a custom-built JSON document store. Clients store objects in \xdb{} by
allocating them in the shared memory and passing their references to the
database along with a key. \xdb{} then takes ownership of the object and
associates the object with the key. The clients can read or write to this object
by sending \xdb{} a read request with the corresponding key.  In return, it
receives pointer to the in-memory data structure that holds the data.

To evaluate \xdb{}, we first populate it with 100k JSON documents using the
NoBench load generator~\cite{nobench} (labeled ``build'' in the figures) and
then issue 1000 JSON search queries to the database (labeled ``search'' in the
figures).

\reffig{fig:cooldb-perf} shows the total runtime of the two operations for the
three versions of \xname{} (CXL, RDMA, and Secure), ZhangRPC, and eRPC. Overall,
\xname{} outperforms all other RPC frameworks when running over CXL, including
Zhang RPC. However, it slows down considerably when running over RDMA during the
build phase, as the shared memory needs to copy multiple pages back and
forth. Moreover, as \xname{} does not need to serialize the dataset or the
queries, it considerably outperforms eRPC for the search operation.

\begin{figure}
  \includegraphics[width=\linewidth]{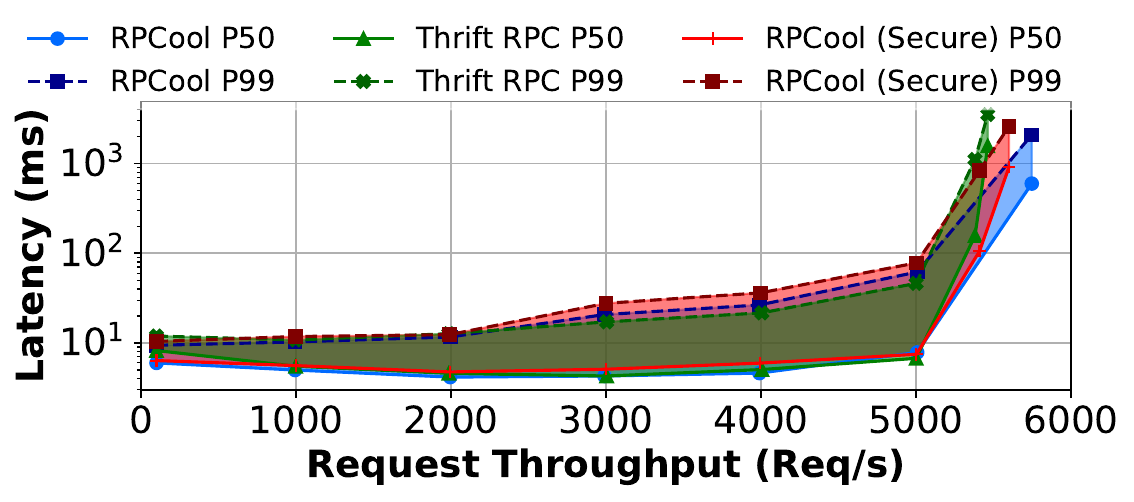}
  \caption{DeathStarBench SocialNetwork Benchmark media and P99 latencies using
    ThriftRPC and \xname{}.}
  \label{fig:deathstarbench-tput-vs-lat}
  \vspace{-0.15cm}
\end{figure}

\begin{figure}
  \includegraphics[width=\linewidth]{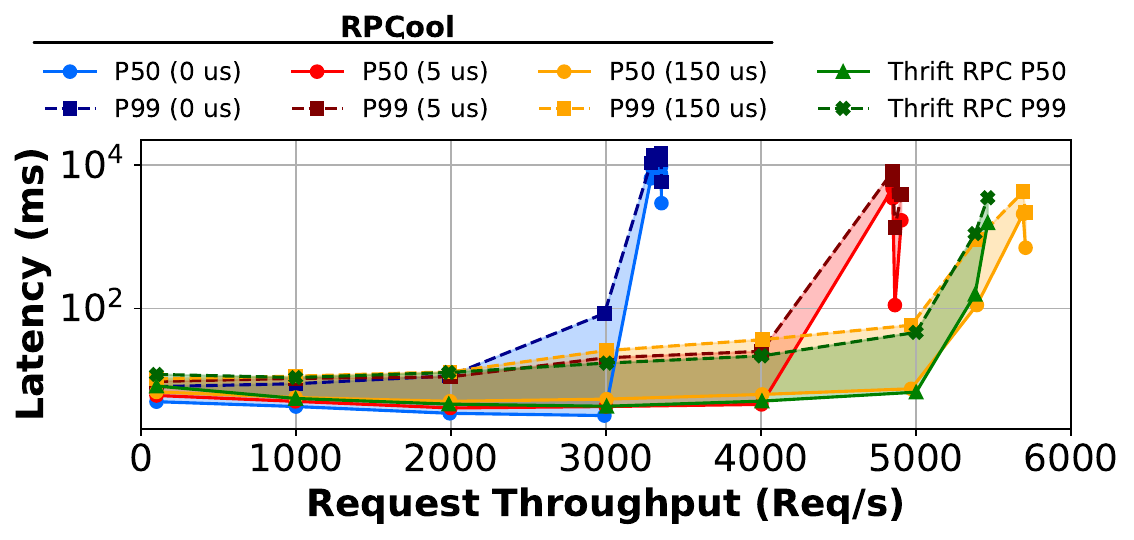}
  \caption{Throughput-latency tradeoff with varying busy wait sleep times in \xname{}.}
  \label{fig:deathstarbench-tput-vs-lat-var-lat}
  \vspace{-0.15cm}
\end{figure}

\paragraph{DeathStarBench's Social Network} We evaluate \xname{} using the
Social Network benchmark from DeathStarBench~\cite{deathstarbench}, which models
a social networking website. In our evaluation, we replace all ThriftRPC calls
among microservices with \xname{} on our emulated CXL platform.  However, as
DeathStarBench spawns multiple new threads for each request, it contends for the
kernel page table lock with \xname{}'s \seal{} and \release{} calls. To address
the issue, we modify the benchmark to use a thread pool instead of creating new
threads for each request in both the ThriftRPC and \xname{} versions.
Additionally, we replaced MongoDB with its RPCool version.

We run DeathStarBench's benchmark that creates user posts under a range of
offered loads and measure the median and P99 latency, as shown in
\reffig{fig:deathstarbench-tput-vs-lat}. The experiment is run for 30~seconds
for each data point. The results demonstrate that \xname{} (both secure and
insecure versions) and ThriftRPC show similar performance, with \xname{}'s peak
throughput surpassing that of ThriftRPC.

To understand why RPCool performs comparably to Thrift RPC, we looked at where a
request spends its time using DeathStarBench’s built-in tracing. We found that,
on average, about 66\% of a request’s critical path latency is spent in
databases and Nginx, suggesting that DeathStarBench's performance is largely
bound by database updates and Nginx.

Further, \reffig{fig:deathstarbench-tput-vs-lat-var-lat} presents benchmark
results with 0~\us{}, 5~\us{}, and 150~\us{} sleep between busy-wait iterations
(\refsec{subsec:busy-wait}).  Not sleeping between iterations results in the
best latencies but limited throughput as busy waiting consumes a significant
amount of CPU.  Conversely, 150~\us{} sleep duration results in higher tail
latencies but achieves higher peak throughput.  Thus, \xname{} offers the
flexibility to balance latency and throughput according to specific needs.


\widowpenalties 1 0

\section{Related Work}
\label{sec:related}

Some prior works have proposed using RPCs over distributed shared memory.
Similar to \xname{}, Wang et al.~\cite{wang2021in} describe RPCs with references
to objects over distributed shared memory. However, since they focus on
data-intensive applications, they propose immutable RPC arguments and return
values and require trust among the applications.  Some works also optimize the
RPC boundary; Nu~\cite{ruan2023nu} optimizes microservices by reimagining how
they are composed. Nu breaks down web applications into proclets that share the
same address space among multiple hosts and uses optimized RPCs for
communication among them. When proclets are placed on the same machine, they
make local function calls, and traditional RPCs otherwise. However, in both
cases, proclets need to copy the arguments to the receiver and require mutual
trust. Lu et al.~\cite{lu2024serialization} improve the performance of
serverless functions by implementing \texttt{rmap()}, allowing serverless
functions to map remote memory, thus avoiding serialization. However,
\texttt{rmap()} requires mutual trust between the sender and the receiver.

Numerous prior studies have explored optimizing the performance of RPC
frameworks using RDMA, but they all require serialization and compression,
adding performance overheads.  HatRPC~\cite{hatrpc} uses code hints to optimize
Thrift RPC and enables RDMA verbs-based communication, while DaRPC~\cite{darpc}
implements an optimized RDMA-based custom RPC framework. Kalia et
al.~\cite{erpc} propose a highly efficient RDMA-based RPC framework called eRPC
that outperforms traditional TCP-based RPCs in latency and throughput.
\ignore{\textsc{Flock}~\cite{monga2021birds} improves RDMA performance in high
  fan-out and fan-in RDMA-based system by dynamic connection sharing among
  threads.} Chen et al.~\cite{chen2023remote} avoid the overhead of sidecars
used in RPC deployment by implementing serialization and sidecar policies as a
system service. Sidecars are proxy processes that run alongside the main
application for policy enforcement, logging, etc., without modifying the
application.

Zhang et al.~\cite{zhang2023partial} present a memory management system for
CXL-based shared memory. Their implementation provides failure resilience
against memory leaks without significant performance overheads. In addition to
failure resiliency, Zhang et al. also propose CXL-based shared memory RPCs,
which we refer to as Zhang RPC. However, Zhang RPC performs significantly slower
compared to \xname{} (\reftab{tab:noop-latency}), does not scale beyond a rack,
and requires mutual trust among applications.  Another CXL-based RPC framework,
DmRPC~\cite{zhang2024dmrpc} supports RPCs over CXL, however, it requires
serialization and mutual trust among processes.

Some works have combined CXL-based shared memory with other communication
protocols. CXL over Ethernet~\cite{cxl-over-ethernet} uses a host-attached CXL
FPGA to transmit CXL.mem requests over Ethernet, enabling host-transparent
Ethernet-based remote memory. Rcmp~\cite{rcmp} overcomes the limited scalability
of CXL-based shared memory by extending it using RDMA. However, similar to
\texttt{rmap()}, it requires applications to mutually trust each other.

Simpson et al.~\cite{simpson2020securing} explore the security challenges of
deploying RDMA in the datacenter. The challenges listed in their work, e.g.,
unauditable writes and concurrency problems, are shared by \xname{} and other
RDMA-based systems alike. Chang et al.~\cite{chang1998security} discuss the
performance overhead of untrusted senders, as the receiver would need to
validate the received pointers and data types. Similar to \xname{}, for
single-machine communication, Chang et al. propose zero-copy RPCs by directly
reading the sender's buffer in trusted environments. Schmidt et
al.~\cite{schmidt1996using} propose a shared memory read-mostly RPC design where
the clients have unrestricted read access to a server's data over shared memory
but make protected and expensive RPCs to update it. Further, since the clients
cannot hold locks in the shared memory, they implement a multi-version
concurrency control to allow updates to the data while clients are reading
them. Schmidt et al.'s solution is orthogonal to \xname{} and can be combined
with it by ensuring read-only permissions for channels in clients and exporting
separate secure channels for updates. Finally, ERIM~\cite{vahldiek2019erim} uses
MPK to isolate sensitive data and to restrict arbitrary code from accessing
protected regions. However, unlike \xname{} which confines accesses to a shared
memory region while processing an RPC, ERIM uses MPK for protecting sensitive
data from malicious components.

Several prior works, including FaRM~\cite{dragojevic2014farm},
RAMCloud~\cite{ramcloud}, \ignore{Grappa~\cite{nelson2015latency},}
Carbink~\cite{zhou2022carbink}, Hydra~\cite{lee2022hydra}, and
AIFM~\cite{ruan2020aifm} enable distributed shared memory and support varying
levels of failure resiliency. However, they require application support for
reads and writes and often use non-standard pointers, breaking compatibility
with legacy code and adding programming overhead. In contrast, \xname{} supports
the same \load{}/\store{} semantics for CXL- and RDMA-based shared memory.
Further, while \xname{}'s RDMA fallback does not implement erasure coding, its
design does not preclude such features.


\section{Conclusion}
\label{sec:conclude}
This work presents \xname{}, a fast, scalable, and secure shared memory RPC
framework for the CXL-enabled world of rack-scale coherent shared memory.  While
shared memory RPCs are fast, they are vulnerable to invalid/wild pointers and
the sender concurrently modifying data with the receiver. Furthermore, CXL is
limited to a rack (e.g., up to 32 nodes).

\xname{} addresses these challenges by preventing the sender from modifying
in-flight data using seals, processing shared data in a low-overhead sandbox to
avoid invalid or wild pointers, and automatically falling back to RDMA for
scaling beyond a rack. Overall, RPCool either performs comparably or outperforms
traditional RPC techniques.

\bibliography{common,paper}

\begin{thebibliography}{10}

\bibitem{amit2020don}
Nadav Amit, Amy Tai, and Michael Wei.
\newblock Don't shoot down {TLB} shootdowns!
\newblock In {\em Proceedings of the Fifteenth European Conference on Computer
  Systems}, pages 1--14, 2020.

\bibitem{birrell1984implementing}
Andrew~D Birrell and Bruce~Jay Nelson.
\newblock Implementing remote procedure calls.
\newblock {\em ACM Transactions on Computer Systems (TOCS)}, 2(1):39--59, 1984.

\bibitem{chang1998security}
Chi-Chao Chang, Grzegorz Czajkowski, Chris Hawblitzel, Deyu Hu, and Thorsten
  von Eicken.
\newblock Security versus performance tradeoffs in rpc implementations for safe
  language systems.
\newblock In {\em Proceedings of the 8th ACM SIGOPS European Workshop on
  Support for Composing Distributed Applications}, EW 8, page 158–161.
  Association for Computing Machinery, 1998.

\bibitem{nobench}
Craig Chasseur, Yinan Li, and Jignesh~M. Patel.
\newblock Enabling {JSON} document stores in relational systems.
\newblock In {\em International Workshop on the Web and Databases}, June 2013.

\bibitem{chen2023remote}
Jingrong Chen, Yongji Wu, Shihan Lin, Yechen Xu, Xinhao Kong, Thomas Anderson,
  Matthew Lentz, Xiaowei Yang, and Danyang Zhuo.
\newblock Remote procedure call as a managed system service.
\newblock In {\em 20th {USENIX} Symposium on Networked Systems Design and
  Implementation (NSDI '23)}, pages 141--159, 2023.

\bibitem{ipoib}
Jerry Chu and Vivek Kashyap.
\newblock Transmission of {IP} over {InfiniBand} ({IPoIB}).
\newblock \url{https://www.rfc-editor.org/rfc/rfc4391.txt}, April 2006.

\bibitem{ycsb-scan}
cimballihw.
\newblock {memcached SCAN always fail \#668}.
\newblock GitHub issue, March 2016.
\newblock GitHub repository:
  \url{https://github.com/brianfrankcooper/YCSB/issues/668}.

\bibitem{ycsb}
Brian~F. Cooper, Adam Silberstein, Erwin Tam, Raghu Ramakrishnan, and Russell
  Sears.
\newblock {Benchmarking Cloud Serving Systems with YCSB}.
\newblock In {\em Proceedings of the 1st ACM Symposium on Cloud Computing},
  SoCC '10, pages 143--154. Association for Computing Machinery, 2010.

\bibitem{dragojevic2014farm}
Aleksandar Dragojevi{\'c}, Dushyanth Narayanan, Miguel Castro, and Orion
  Hodson.
\newblock {FaRM}: Fast remote memory.
\newblock In {\em 11th USENIX Symposium on Networked Systems Design and
  Implementation (NSDI '14)}, pages 401--414, Seattle, WA, April 2014. USENIX
  Association.

\bibitem{deathstarbench}
Yu~Gan, Yanqi Zhang, Dailun Cheng, Ankitha Shetty, Priyal Rathi, Nayan Katarki,
  Ariana Bruno, Justin Hu, Brian Ritchken, Brendon Jackson, et~al.
\newblock An open-source benchmark suite for microservices and their
  hardware-software implications for cloud \& edge systems.
\newblock In {\em Proceedings of the Twenty-Fourth International Conference on
  Architectural Support for Programming Languages and Operating Systems}, pages
  3--18, 2019.

\bibitem{boost.interprocess}
Ion Gaztanaga.
\newblock {\em Boost 1.79.0 Documentation}, chapter {Boost.Interprocess}.
\newblock 2022.

\bibitem{grpc}
{Google Inc.}
\newblock {gRPC}, 2021.
\newblock \url{https://grpc.io/}. Accessed: 2023-02-21.

\bibitem{dax-cxl-lpc}
John Groves.
\newblock Shared {CXL} 3 memory: what will be required?
\newblock Linux Plumbers Conference, November 2023.
\newblock \url{https://lpc.events/event/17/contributions/1455/}.

\bibitem{erpc}
Anuj Kalia, Michael Kaminsky, and David Andersen.
\newblock Datacenter {RPCs} can be general and fast.
\newblock In {\em 16th {USENIX} Symposium on Networked Systems Design and
  Implementation (NSDI '19)}, pages 1--16, 2019.

\bibitem{argodsm}
Stefanos Kaxiras, David Klaftenegger, Magnus Norgren, Alberto Ros, and
  Konstantinos Sagonas.
\newblock Turning centralized coherence and distributed critical-section
  execution on their head: A new approach for scalable distributed shared
  memory.
\newblock In {\em Proceedings of the 24th International Symposium on
  High-Performance Parallel and Distributed Computing}, pages 3--14, 2015.

\bibitem{lee2022hydra}
Youngmoon Lee, Hasan Al~Maruf, Mosharaf Chowdhury, Asaf Cidon, and Kang~G Shin.
\newblock Hydra: Resilient and highly available remote memory.
\newblock In {\em 20th USENIX Conference on File and Storage Technologies (FAST
  '22)}, pages 181--198, 2022.

\bibitem{hatrpc}
Tianxi Li, Haiyang Shi, and Xiaoyi Lu.
\newblock {HatRPC}: hint-accelerated {Thrift} {RPC} over {RDMA}.
\newblock In {\em Proceedings of the International Conference for High
  Performance Computing, Networking, Storage and Analysis}, SC '21. Association
  for Computing Machinery, 2021.

\bibitem{lu2024serialization}
Fangming Lu, Xingda Wei, Zhuobin Huang, Rong Chen, Minyu Wu, and Haibo Chen.
\newblock Serialization/deserialization-free state transfer in serverless
  workflows.
\newblock In {\em Proceedings of the Nineteenth European Conference on Computer
  Systems}, pages 132--147, 2024.

\bibitem{mahar2024puddles}
Suyash Mahar, Mingyao Shen, TJ~Smith, Joseph Izraelevitz, and Steven Swanson.
\newblock Puddles: Application-independent recovery and location-independent
  data for persistent memory.
\newblock In {\em Proceedings of the Nineteenth European Conference on Computer
  Systems}, EuroSys '24, page 575–589. Association for Computing Machinery,
  2024.

\bibitem{memcached}
Memcached.
\newblock http://memcached.org/.

\bibitem{mongodb}
{MongoDB, Inc.}
\newblock {MongoDB}, 2017.
\newblock \url{https://www.mongodb.com}.

\bibitem{cxl-switch}
James Morra.
\newblock {CXL} switch {SoC} unlocks more memory for {AI}, 2023.
\newblock Retrieved from
  \url{https://www.electronicdesign.com/technologies/embedded/article/21272132/electronic-design-cxl-switch-soc-unlocks-more-memory-for-ai}.

\bibitem{ramcloud}
John Ousterhout, Arjun Gopalan, Ashish Gupta, Ankita Kejriwal, Collin Lee,
  Behnam Montazeri, Diego Ongaro, Seo~Jin Park, Henry Qin, Mendel Rosenblum,
  Stephen Rumble, Ryan Stutsman, and Stephen Yang.
\newblock {The RAMCloud Storage System}.
\newblock {\em ACM Trans. Comput. Syst.}, 33(3):7:1--7:55, August 2015.

\bibitem{libmpk}
Soyeon Park, Sangho Lee, Wen Xu, Hyungon Moon, and Taesoo Kim.
\newblock libmpk: Software abstraction for {Intel} memory protection keys
  ({Intel} {MPK}).
\newblock In {\em 2019 {USENIX} Annual Technical Conference (USENIX ATC '19)},
  pages 241--254, 2019.

\bibitem{boost.pfr}
Antony Polukhin.
\newblock {\em Boost 1.84.0 Documentation}, chapter 26. {Boost.PFR 2.2}.
\newblock 2023.

\bibitem{ruan2023nu}
Zhenyuan Ruan, Seo~Jin Park, Marcos~K Aguilera, Adam Belay, and Malte
  Schwarzkopf.
\newblock Nu: Achieving {Microsecond-Scale} resource fungibility with logical
  processes.
\newblock In {\em 20th {USENIX} Symposium on Networked Systems Design and
  Implementation (NSDI '23)}, pages 1409--1427, 2023.

\bibitem{ruan2020aifm}
Zhenyuan Ruan, Malte Schwarzkopf, Marcos~K Aguilera, and Adam Belay.
\newblock {AIFM}: {High-Performance},{Application-Integrated} far memory.
\newblock In {\em 14th USENIX Symposium on Operating Systems Design and
  Implementation (OSDI '20)}, pages 315--332, 2020.

\bibitem{schmidt1996using}
Rene~W Schmidt, Henry~M Levy, and Jeffrey~S Chase.
\newblock Using shared memory for read-mostly {RPC} services.
\newblock In {\em Proceedings of HICSS-29: 29th Hawaii International Conference
  on System Sciences}, volume~1, pages 141--149. IEEE, 1996.

\bibitem{google-rpc-study}
Korakit Seemakhupt, Brent~E Stephens, Samira Khan, Sihang Liu, Hassan Wassel,
  Soheil~Hassas Yeganeh, Alex~C Snoeren, Arvind Krishnamurthy, David~E Culler,
  and Henry~M Levy.
\newblock A cloud-scale characterization of remote procedure calls.
\newblock In {\em Proceedings of the 29th Symposium on Operating Systems
  Principles}, pages 498--514, 2023.

\bibitem{cxl}
Debendra~Das Sharma.
\newblock {Compute Express Link}{\textregistered}: An open industry-standard
  interconnect enabling heterogeneous data-centric computing.
\newblock In {\em 2022 IEEE Symposium on High-Performance Interconnects
  (HOTI)}, pages 5--12. IEEE, 2022.

\bibitem{simpson2020securing}
Anna~Kornfeld Simpson, Adriana Szekeres, Jacob Nelson, and Irene Zhang.
\newblock Securing {RDMA} for {High-Performance} datacenter storage systems.
\newblock In {\em 12th {USENIX} Workshop on Hot Topics in Cloud Computing
  (HotCloud '20)}. {USENIX} Association, July 2020.

\bibitem{thriftrpc}
Mark Slee, Aditya Agarwal, and Marc Kwiatkowski.
\newblock Thrift: Scalable cross-language services implementation.
\newblock {\em Facebook white paper}, 5(8):127, 2007.

\bibitem{darpc}
Patrick Stuedi, Animesh Trivedi, Bernard Metzler, and Jonas Pfefferle.
\newblock {DaRPC}: Data center {RPC}.
\newblock In {\em Proceedings of the ACM Symposium on Cloud Computing}, pages
  1--13, 2014.

\bibitem{sung2020intra}
Mincheol Sung, Pierre Olivier, Stefan Lankes, and Binoy Ravindran.
\newblock Intra-unikernel isolation with {Intel} memory protection keys.
\newblock In {\em Proceedings of the 16th ACM SIGPLAN/SIGOPS International
  Conference on Virtual Execution Environments}, pages 143--156, 2020.

\bibitem{vahldiek2019erim}
Anjo Vahldiek-Oberwagner, Eslam Elnikety, Nuno~O Duarte, Michael Sammler, Peter
  Druschel, and Deepak Garg.
\newblock {ERIM}: Secure, efficient in-process isolation with protection keys
  ({MPK}).
\newblock In {\em 28th USENIX Security Symposium (USENIX Security 19)}, pages
  1221--1238, 2019.

\bibitem{cxl-over-ethernet}
Chenjiu Wang, Ke~He, Ruiqi Fan, Xiaonan Wang, Wei Wang, and Qinfen Hao.
\newblock {CXL} over {Ethernet}: A novel {FPGA}-based memory disaggregation
  design in data centers.
\newblock In {\em 2023 IEEE 31st Annual International Symposium on
  Field-Programmable Custom Computing Machines (FCCM)}, pages 75--82. IEEE,
  2023.

\bibitem{wang2021in}
Stephanie Wang, Benjamin Hindman, and Ion Stoica.
\newblock In reference to {RPC}: it's time to add distributed memory.
\newblock In {\em Proceedings of the Workshop on Hot Topics in Operating
  Systems}, HotOS '21, page 191–198, New York, NY, USA, 2021. Association for
  Computing Machinery.

\bibitem{rcmp}
Zhonghua Wang, Yixing Guo, Kai Lu, Jiguang Wan, Daohui Wang, Ting Yao, and
  Huatao Wu.
\newblock Rcmp: Reconstructing {RDMA}-based memory disaggregation via {CXL}.
\newblock {\em ACM Transactions on Architecture and Code Optimization},
  21(1):1--26, 2024.

\bibitem{zhang2024dmrpc}
Jie Zhang, Xuzheng Chen, Yin Zhang, and Zeke Wang.
\newblock {DmRPC: Disaggregated Memory-aware Datacenter RPC for Data-intensive
  Applications}.
\newblock In {\em 40th IEEE International Conference on Data Engineering
  (ICDE)}, Utrecht, Netherlands, May 13-17 2024. IEEE, IEEE.

\bibitem{zhang2023partial}
Mingxing Zhang, Teng Ma, Jinqi Hua, Zheng Liu, Kang Chen, Ning Ding, Fan Du,
  Jinlei Jiang, Tao Ma, and Yongwei Wu.
\newblock Partial failure resilient memory management system for ({CXL}-based)
  distributed shared memory.
\newblock In {\em Proceedings of the 29th Symposium on Operating Systems
  Principles}, SOSP'23, page 658–674. Association for Computing Machinery,
  2023.

\bibitem{zhou2022carbink}
Yang Zhou, Hassan~MG Wassel, Sihang Liu, Jiaqi Gao, James Mickens, Minlan Yu,
  Chris Kennelly, Paul Turner, David~E Culler, Henry~M Levy, et~al.
\newblock Carbink: {Fault-Tolerant} far memory.
\newblock In {\em 16th USENIX Symposium on Operating Systems Design and
  Implementation (OSDI '22)}, pages 55--71, 2022.

\end{thebibliography}
\bibliographystyle{plain}

\end{document}